\documentclass{aa}

\usepackage{xspace}
\usepackage{amssymb}
\usepackage{amsmath}
\usepackage{amsfonts}
\usepackage[dvips]{graphicx}
\usepackage{subfigure}

\newcommand{\teff}{$T_\mathrm{eff}$\xspace} 
\newcommand{\lgg}{$\log{g}$\xspace}
\newcommand{\vrad}{$V_\mathrm{rad}$\xspace}
\newcommand{\vsini}{$V\sin{i}$\xspace}
\newcommand{\vmic}{$\xi_\mathrm{micro}$\xspace}
\newcommand{\kms}{$\mathrm{km~s^{-1}}$\xspace}
\newcommand{\msol}{M$_\odot$\xspace}

\begin{document}

\title{Automated Spectroscopic abundances of A and F-type stars\\ using echelle
  spectrographs}

\subtitle{II. Abundances of 140 A-F stars from ELODIE\thanks{Based on
    observations collected at the 1.93\,m
    telescope at the Observatoire de Haute-Provence (St-Michel
    l'Observatoire, France)} and CORALIE\thanks{Based on observations
    collected at the Swiss 1.2\,m Leonard Euler telescopes at the European
    Southern Observatory (La Silla, Chile)} \thanks{Tables 5 and 6 are
    only available in electronic form at the 
    CDS via anonymous ftp to cdsarc.u.strasbg.fr (130.79.128.5) or via
    http://cdsweb.u-strasbg.fr/Abstract.html}}

\author{D. Erspamer \and P. North}

\offprints{P. North \\
\email{Pierre.North@obs.unige.ch}}

\institute{Institut d'astronomie de l'Universit\'e de Lausanne, 
              CH-1290 Chavannes-des-Bois, Switzerland}

\date{Received / Accepted }

\abstract{Using the method presented in Erspamer \& North
  (\cite{erspamer}, Paper I hereafter), detailed abundances of 140
stars are presented. The uncertainties characteristic of this method are
presented and discussed. In particular, we show that for a 
S/N ratio higher than 200, the method is applicable to stars with a rotational
velocity as high as 200 \kms. 
There is no correlation between abundances and \vsini, except a spurious one for Sr, Sc
and Na which we explain 
by the small number of lines of these elements combined with a locally 
biased continuum.
Metallic giants (Hauck, \cite{hauck}) show larger abundances than
normal giants for at least 8 elements: Al, Ca, Ti, Cr, Mn, Fe, Ni and Ba. 
The anticorrelation for Na, Mg, Si, Ca, Fe and Ni with \vsini
suggested by Varenne and Monier (\cite{varenne99}) is not
confirmed. The predictions of the
Montr\'eal models (e.g. Richard et al., \cite{richard01}) are not fulfilled in general. 
However a  correlation between  $\left[\frac{Fe}{H}\right]$
and \lgg is found for stars of 1.8 to 2.0 \msol. Various possible
causes are discussed, but the physical reality of this correlation
seems inescapable.
}

\titlerunning{Abundances of 140 A-F stars with ELODIE and CORALIE}

\maketitle

\section{Introduction}
A-F type stars present a large number of chemical peculiarities: Ap, Am,
$\lambda$~Bootis, $\rho$~Puppis (see Gray \& Garrison
\cite{gray89}, for a discussion of the $\delta$~Del stars and a
definition of the $\rho$~Pup stars). Radiative diffusion is 
generally invoked most of these peculiarities, and the theory of this mechanism,
developed mostly by Michaud and collaborators (Michaud \cite{michaud}) seems
indeed largely successful, even though some adjustments seem to remain
necessary. For instance, stellar wind was added to radiative diffusion in order
to make more realistic predictions of abundances of Am stars (Michaud
et al. \cite{michaud83}).

A little explored category of chemically peculiar stars in this range of
spectral types is that of metallic giants, discovered by Hauck (\cite{hauck})
on the basis of textsc{Geneva} photometry and further studied by Berthet
(\cite{berthet90, berthet91}) using spectroscopy. They are A-F stars
leaving the main sequence and having the Geneva photometric parameter
$\Delta\mathrm{m_2} \geq 0.013$ (see Hauck, \cite{hauck73} for the definition
of $\Delta\mathrm{m_2}$). Berthet (\cite{berthet90, berthet91}) confirmed that
these stars have metal abundances higher than solar ones, as suggested by their
large $\Delta\mathrm{m_2}$ values. However, he analyzed only metallic
giants and 
relied on published abundances of normal giants for comparison. Berthet found
that the abundance pattern of metallic giants is similar to that of Am stars,
except for calcium and scandium, which are notoriously underabundant in the
latter, while they are roughly normal or slightly overabundant in the former.
On this basis, Berthet (\cite{berthet92}) proposed that metallic
giants might be 
evolved Am stars, an
idea theoretically justified by the following scenario: in the beginning of the
life of an Am star, Ca and Sc sink and accumulate below the line forming region
because of their too small radiative acceleration. During stellar evolution on
the main sequence, however, the effective temperature decreases and the outer
convection zone develops and thickens, until it reaches the depth where Ca and
Sc have accumulated and dredges them up, so that their surface abundances
becomes normal again. K\"unzli \& North (\cite{kunzli98}) had challenged this
scenario, arguing, in particular, that the \vsini distribution of metallic
giants is not compatible with that of Am stars, especially evolved ones. The
fact that metallic giants are more evolved and more massive than normal ones
(North et al. \cite{north98}) does not fit Berthet's scenario either.

The initial purpose of this work was to focus our attention on
metallic giants by comparing their abundances with those of normal giants, in
order to better define their characteristics and determine how far they form a
well defined spectroscopic group. However, since Berthet's scenario remains
unsatisfactory, the question of their origin remains open, so we decided to
examine these stars in a more general context and try to understand what are
their progenitors. So we judged important
to study not only giant A-F stars (normal and metallic) but also their 
possible progenitors, i.e. normal A-F type stars of similar masses on the main
sequence. To this purpose, we
chose to observe a sample of A-F stars that cover the HR diagram as 
homogeneously as possible, in order to map the abundance pattern in the
whole region of interest. In addition to the homogeneous coverage of the HR
diagram, it is important to ensure also the homogeneity of the method of
spectral analysis. In the latter sense, homogeneity is crucial, since
it is very delicate to 
compare abundance analyses done by different authors. There is always
at least one assumption that differs. The most common problem is the
choice of atmospheric parameters, \teff and \lgg. Indeed, different
techniques, using different photometric systems  
or spectroscopy, are used to derive atmospheric
parameters. Abundance differences arising from these techniques are
generally not very
important since they remain within error boxes, but they significantly
increase the scatter in the results. Another common difference
is the choice of atmosphere model theory. In the field of A-F stars,
ATLAS9 models from Kurucz (\cite{kurucz1}) are the rule. However, these
models exist with and without overshooting, and both are used depending on the
authors. Finally, methods to derive the abundances differ. Some
people use equivalent widths while others prefer synthetic spectra
adjustments. The oscillator strengths and other line parameters may
come from different sources and these sources may vary from one element to 
the next. The adopted solar iron abundance may also differ from one
author to the other. Therefore, even for iron, a large scatter is
found. For example, even for a very bright star such as Procyon, [Fe/H]
values vary from -0.18 to 0.05 (Cayrel de Strobel et al. \cite{cayrel}).
All this shows how well justified is the extensive and homogeneous
analysis of a large sample of A-F type stars that we present in this paper.

In Sect.~\ref{sec:obs}, we describe the observations and the reduction
procedure, while 
Sect.~\ref{sec:sample} gives more details on the definition of the
sample. The determination 
of atmospheric parameters is described in Sect.~\ref{fun_param}, and sources of
uncertainties on the derived abundances are discussed in
Sect.~\ref{sec:uncert}. The results
are presented in Sect.~\ref{sec:res} and discussed in Sect.~\ref{sec:disc}.

\section{Observations and data reduction}\label{sec:obs}

The spectra used in this work were obtained with two almost identical
spectrographs, ELODIE and CORALIE (see Baranne et al., 
\cite{baranne}, for technical details about ELODIE and Queloz et al., 
\cite{queloz}, for CORALIE). Both spectrographs have been built as a
joint project between the Geneva Observatory and the Observatoire de
Haute-Provence. ELODIE is  attached to the 1.93\,m
telescope of the Observatoire de 
Haute-Provence whereas CORALIE is attached to the Swiss 1.2\,m Leonard 
Euler telescope at the ESO la Silla Observatory. The optical design of 
both spectrographs is the same. The major difference is the better resolving
power of CORALIE (50\,000 compared to 42\,000 for ELODIE) that comes
from a slightly different optical combination at the entrance of the
spectrograph and the use of a 2k$\times$2k CCD camera (against a 1k$\times$1k
for ELODIE) with smaller pixels (15 $\mu$m vs 24 $\mu$m) (Queloz et
al. \cite{queloz99}).  

Two observing runs were held in June and August 1999 in Haute-Provence,
and one in July 2000 in Chile for a total of 23 nights (10 and 13
respectively). Target stars were bright ones, because we wanted a high signal to
noise ratio (S/N $>$ 200). The total number of observed stars is 140,
after elimination of all spectroscopic binaries or stars presenting
strange cross-correlation functions due e.g. to pulsations.

Data reduction was done as presented in Paper I. As the optical design 
of CORALIE is the same as the one of ELODIE, only a few changes in the
parameters used to characterize the CCD were needed to adapt the IRAF
reduction procedure which was initially designed for ELODIE
spectra. However, after order
extraction, the different orders merge perfectly without scaling,
contrary to the case of ELODIE spectra. The
whole spectrum is obtained by a simple concatenation of consecutive
orders with a detection of the overlapping region. As the S/N ratio
is better in the red part of the order than in the blue one, the flux
of the first order is retained for 3/4 of the overlapping region, and 
the flux of the following for the remaining 1/4. The subsequent
operations are done using the same procedure as described in Paper I.

\section{Sample}\label{sec:sample}
The mass range is 1.3 to about 2.8 \msol, encompassing that of most metallic
giants (1.7 to 3 \msol).
As we wanted to observe as many stars as possible during the
observing runs at our disposal, our choice naturally went to the brighter
stars. Two additional requirements were used to define the sample:
availability of \textsc{Geneva} photometry (Golay \cite{G80}; Rufener \& Nicolet
\cite{RN88}; Cramer \cite{C99}) and of \textsc{Hipparcos}
parallaxes (ESA \cite{esa}), in 
order to be able to determine the atmospheric parameters with the
method presented in Sect.~\ref{fun_param}. Known binaries were
exluded too. Finally, we chose to keep only stars with \vsini 
$< 150$ \kms or that have no determination in the bright star catalogue
(Hoffleit \& Jaschek \cite{bsc}).
These were the only restrictions before observing.

During the observing runs, as our main targets were metallic giants, we
focused a little more on them and on their progenitors, i.e. on stars falling
in the mass range 1.7 to 2.5 \msol. Low mass stars had been included in the
sample essentially because the early Montr\'eal models (see Turcotte et
al. (\cite{turcotte98}), Richer et al. (\cite{richer00}) and Richard
et al. (\cite{richard01}) and reference therein) did not consider
masses larger than 1.5 \msol. 

As observations were done with spectrographs designed to get radial
velocities, the cross-correlation peak was routinely available. we used it to
exclude stars presenting a ``dubious'' peak, i.e. an asymmetric one that may
come either from binarity or from pulsation, or a very broad one resulting from
\vsini larger than 200 \kms. 

The final sample is composed of 140 stars, out of which 35 are
giants (18 metallic and 17 normal), 10 are Am and 6 are
Ap. Spectral types are taken from Hauck (\cite{hauck}) for
  stars in common, and from Gray \& Garrison (\cite{gray87, gray89a,
    gray89}) otherwise.
The chemically peculiar Am and Ap stars are included only because their
positions in the HR diagram correspond to the studied region. It is
interesting that without taking care of this detail, the proportion of 
Am and Ap stars, relatively to stars of the main sequence, is
representative. 

\section{Atmospheric  parameters determination}
\label{fun_param}
Atmospheric parameters were determined from \textsc{Geneva}
photometry for \teff and from \textsc{Hipparcos} data (ESA, \cite{esa})
for \lgg. The \texttt{calib} code (K\"unzli et al., \cite{kunzli}) was used 
to derive effective temperature from textsc{Geneva} photometry. Although this
code also provides \lgg, we prefer to determine that value from
\textsc{Hipparcos} data. All stars of the sample are bright and have
an accurate trigonometric parallax measurement by 
\textsc{Hipparcos}. They all have a
parallax larger than 7.4 mas (i.e. are closer than 135 pc except two,
HD 175510 and HD 187764 that are at 166 and 171 pc respectively). Therefore
reddening was neglected \footnote{Although this is completely
    justified for the
121 objects closer than 70~pc on the basis of $uvby\beta$ photometry, a closer
examination requested by the referee shows that the remaining objects are
slightly more reddened on average. In particular, 4 stars have $E(b-y) >
0.020$: HD 115604 ($E(b-y)=0.025$), HD 181333 (0.035), HD 200723 (0.023) and
HD~214441 (0.027), implying that \teff might be underestimated by $\sim 300$~K
in the worst case. Among the 121 nearby stars, 8 have also $E(b-y) > 0.020$,
but they constitute the queue of a fairly gaussian distribution, 6 of them
being closer than 50~pc. The colour excesses were estimated using the usual
calibrations by Crawford (\cite{C75}, \cite{C79}), incorporated in a code
kindly made available by Dr. C. Jordi.} Bolometric luminosity was
calculated using 
parallax, bolometric correction interpolated in Flower's table (\cite{flower})
and $M_{\textrm{bol}}=4.75$ for the Sun. The bolometric luminosity
is then given by 

\[\log{\frac{L}{L_\odot}}= -0.4(M_{\textrm{v}} -4.75+B.C.) \]

The \lgg value was then interpolated in the grids of models of Schaller et al
(\cite{schaller}), Schaerer et al. (\cite{schaerer}), Charbonnel et al. 
(\cite{charbonnel}) and Schaerer et al. (\cite{schaerer1}) according
to the method described by North et al.  (\cite{north}). 
Finally, an ATLAS9 atmosphere model was computed
without overshooting (Castelli et al., \cite{castelli})
with temperature rounded to 25K and \lgg to 0.05 dex using the 
version of ATLAS9 adapted  by
M. Lemke\footnote{http://www.sternwarte.uni-erlangen.de/ftp/michael/ 
atlas-lemke.tgz} for UNIX system. 

\begin{figure}[htbp]
\resizebox{\hsize}{!}{\includegraphics{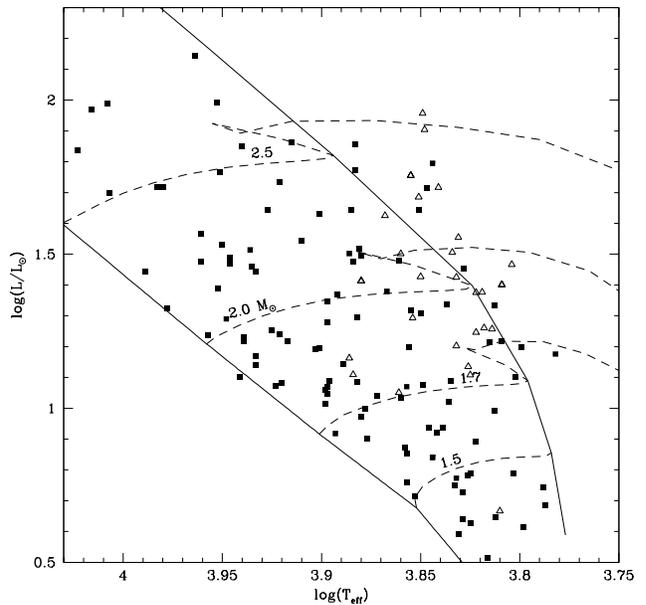}}
\caption{HR diagram of all studied stars with some evolutionary
  tracks (Schaller et al, \cite{schaller}). $\blacksquare$ : stars
  with $\Delta\mathrm{m_2} < 
  0.013$; $\triangle$ : stars with $\Delta\mathrm{m_2} \geq
  0.013$}
\label{diaghr}
\end{figure}

Derived parameters of our sample stars are presented in Table~5
available at CDS. It contains HD, \teff, \lgg, \vrad, \vsini and
\vmic. The last three values are those derived with our method.
Fig.~\ref{diaghr} shows their position in the HR diagram. Stars 
cover the whole main sequence and extend slightly beyond.

\section{Uncertainty estimation}\label{sec:uncert}

\subsection{Errors resulting from wrong atmospheric parameters}
\label{param}
\begin{figure*}[htbp]
  \centering
  \mbox{
        \subfigure[\teff = 6950 ; \lgg = 4.0]{\includegraphics[width=6.cm]{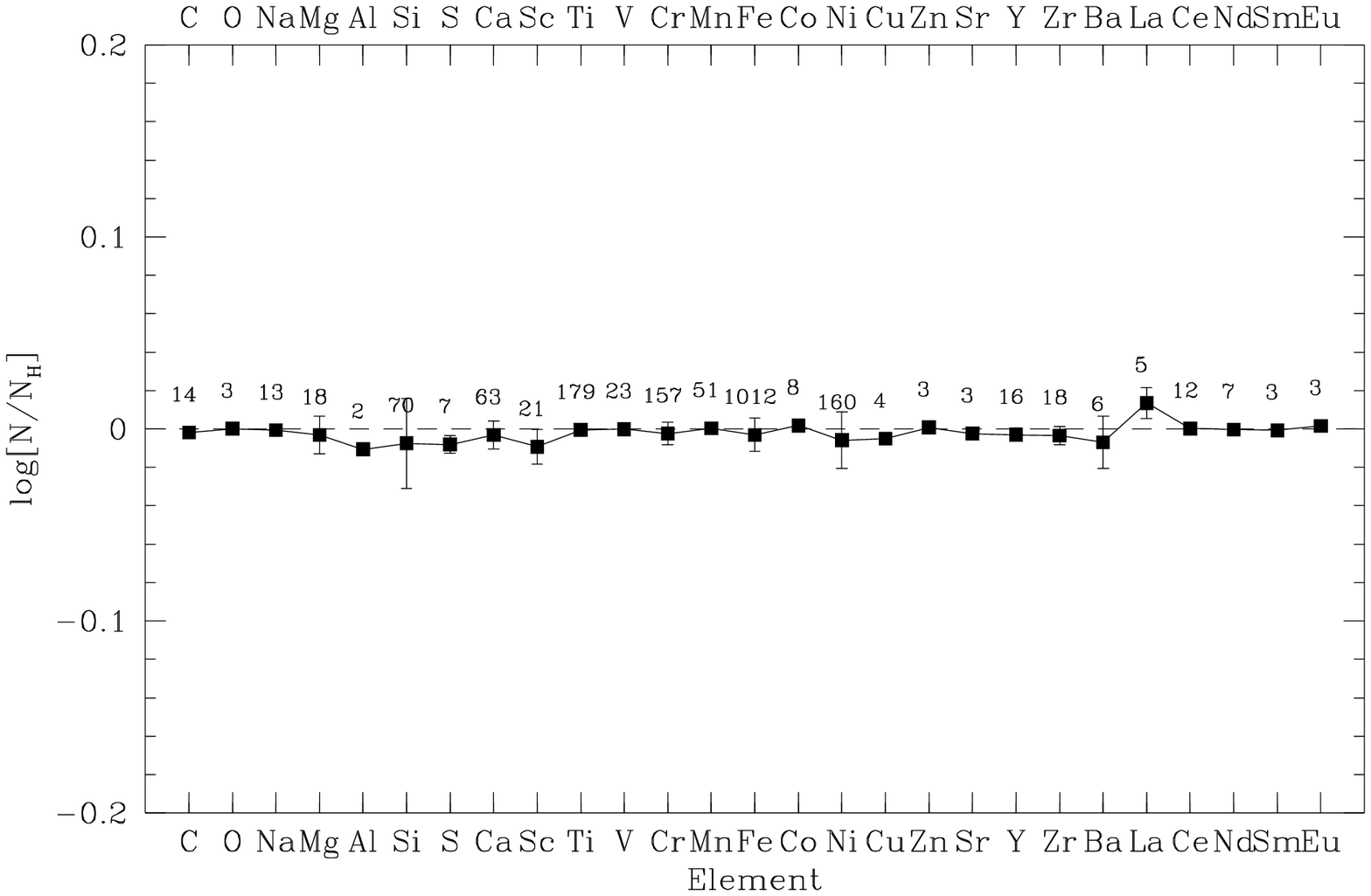}}
        \subfigure[\teff = 6800 ; \lgg = 4.0]{\includegraphics[width=6.cm]{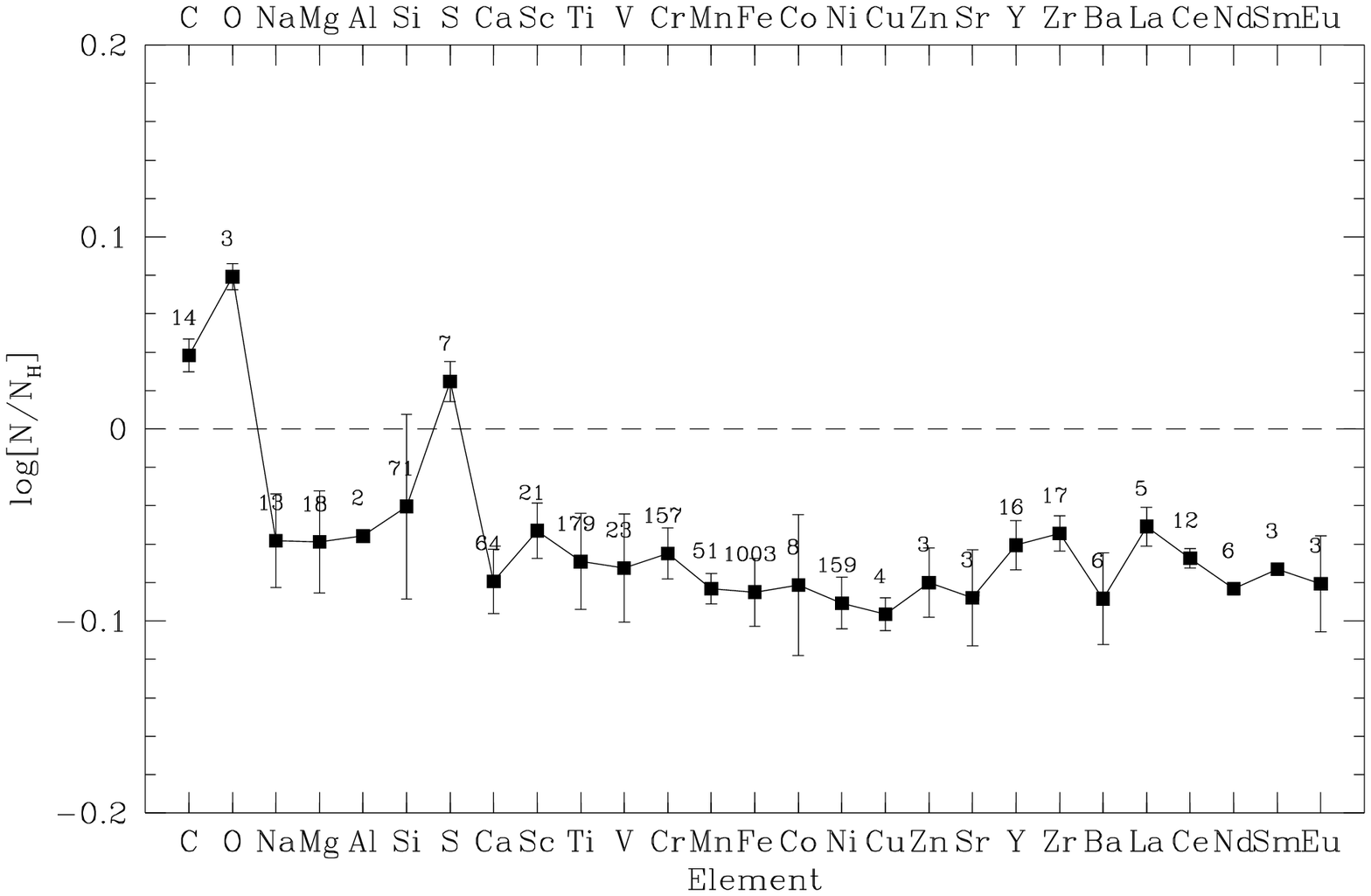}}
        \subfigure[\teff = 7100 ; \lgg = 4.0]{\includegraphics[width=6.cm]{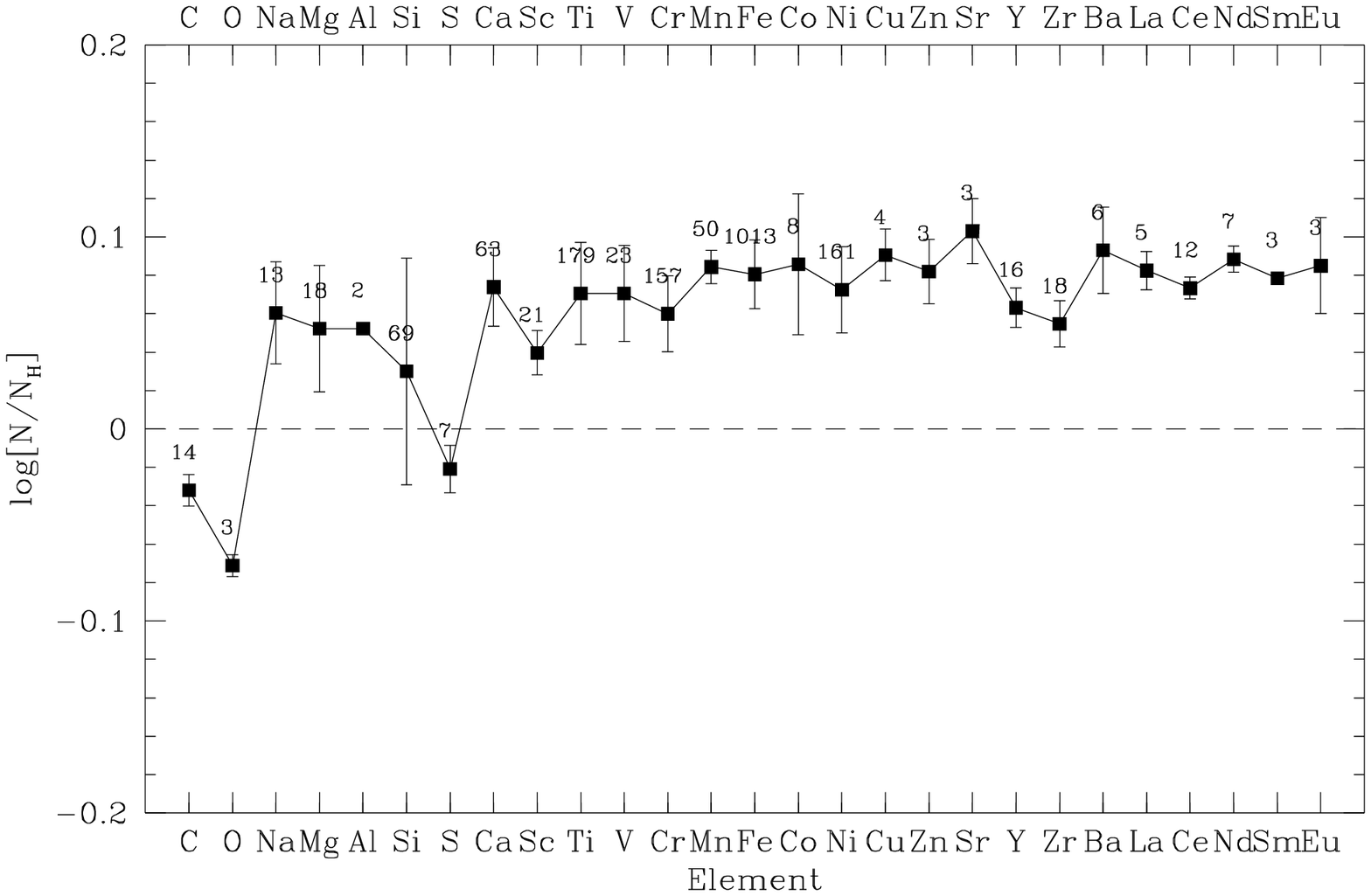}}
  }
  \mbox{
        \subfigure{\includegraphics[width=6.cm]{fig2d.eps}}
        \subfigure[\teff = 6950 ; \lgg = 3.8]{\includegraphics[width=6.cm]{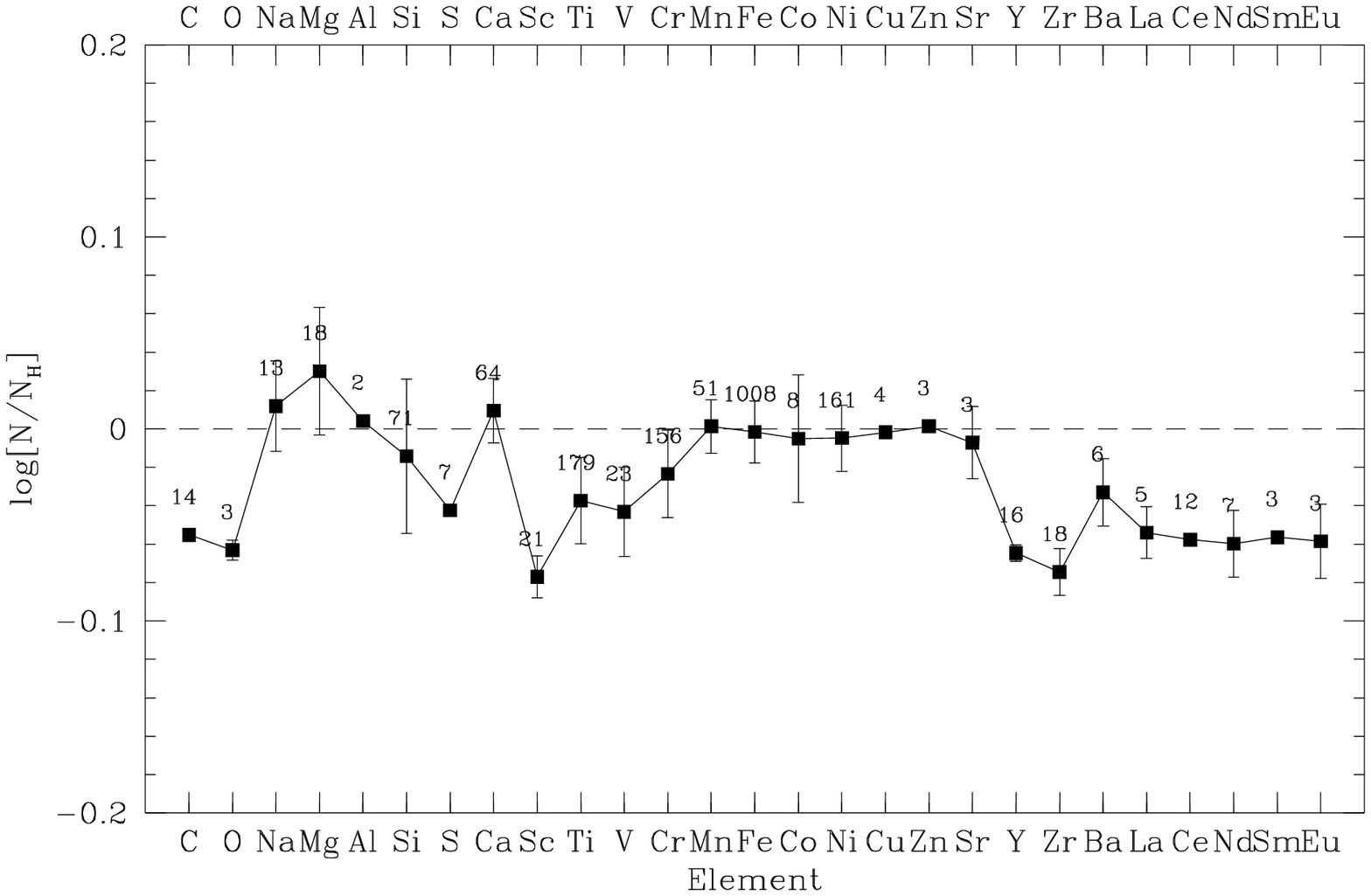}}
        \subfigure[\teff = 6950 ; \lgg = 4.2]{\includegraphics[width=6.cm]{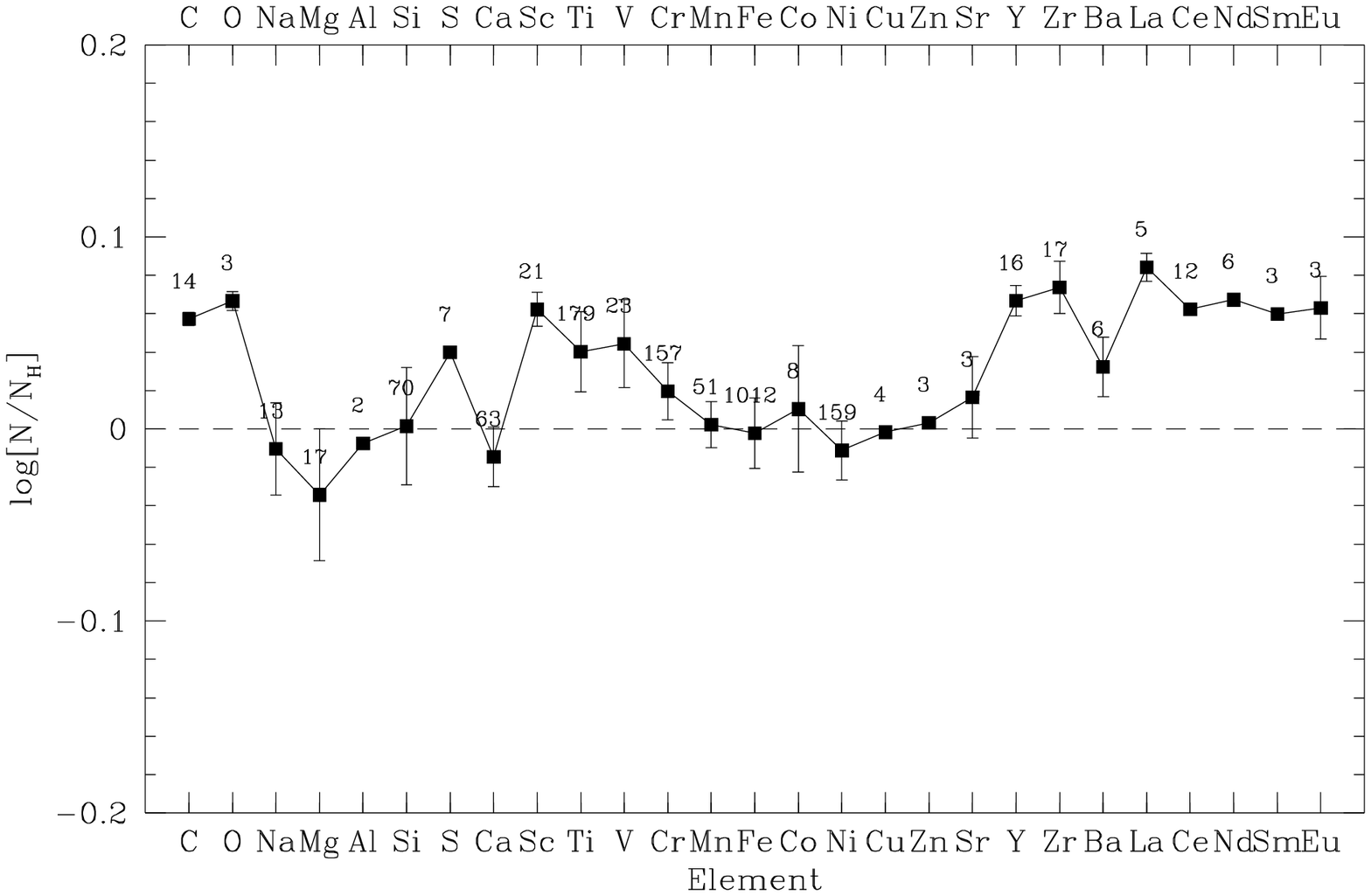}}
  }
  \caption{Abundances with respect to the Sun for different values of
    \teff and \lgg. \textbf{(a)} Abundances obtained with the correct
    atmosphere model. \textbf{(b) to (e)} Abundances obtained with the
    specified parameters. Error bars represent the sigma of the
    average abundance value of the 7 parts of the spectrum (see
    Paper I).}
  \label{error}
\end{figure*}

In order to check the influence of erroneous atmospheric parameters
(\teff $\pm 150 K$, \lgg $\pm 0.2$ dex) on
derived abundances, simulations were done on a synthetic spectrum.
These values do not represent the internal error of the methods used
to derive these parameters, but rather the scatter in the
estimations done with different methods (photometric or spectroscopic).
The synthetic spectrum was computed with a model atmosphere of \teff = 6950 K,
\lgg = 4.0 and solar chemical composition. This model corresponds to a
typical star of our sample 
lying in the middle of both temperature and gravity ranges. 
The other parameters, namely radial, microturbulent and especially rotational
velocities, are \vrad = 0 \kms, \vmic = 1.5 \kms and \vsini= 10 \kms. The
latter was chosen so that the code returns the input abundances
within $\pm 0.01$~dex when abundances are estimated with
the input atmosphere model, the S/N ratio being infinite except for rounding
errors. Thus all variations in derived abundances can be attributed to errors
on \teff or \lgg, even though \vrad, \vmic and \vsini were adjusted, as well
as the abundances. The input abundances are solar.
 The results are presented in Fig.~\ref{error} and listed in
Table~\ref{errortab}

\begin{table}
\caption{Difference between adjusted  and reference abundances in
  hundredth of dex. The last column
is the result of the adjustment done with the correct atmosphere model}
\label{errortab}
\begin{tabular}{c|c|c|c|c|c}
Element & \teff & \teff & \lgg & \lgg & ref\\
        & + 150 K & -150 K & +0.2 dex & -0.2 dex \\
\hline
\vrad  &  0   & 0    & 0   &   0    & 0   \\
\vsini &  0   & 0    & 0   &   0    & 0   \\
\vmic &  0   & 0    & 0   &   0    &  0  \\
C     &  -3  & 4    & 6   &   -6   &  -0 \\
O     &  -7  & 8    & 7   &   -6   &  0  \\
Na    &  6   & -6   & -1  &   1    &  -0 \\
Mg    &  5   & -6   & -3  &   3    &  -0 \\
Al    &  5   & -6   & -1  &   0    &  -1 \\
Si    &  3   & -4   & 0   &   -1   &  -1 \\
S     &  -2  & 2    & 4   &   -4   &  -1 \\
Ca    &  7   & -8   & -1  &   1    &  -0 \\
Sc    &  4   & -5   & 6   &   -8   &  -1 \\
Ti    &  7   & -7   & 4   &   -4   &  -0 \\
V     &  7   & -7   & 4   &   -4   &  -0 \\
Cr    &  6   & -6   & 2   &   -2   &  -0 \\
Mn    &  8   & -8   & 0   &   0    &  0  \\
Fe    &  8   & -8   & -0  &   -0   &  -0 \\
Co    &  9   & -8   & 1   &   -1   &  0  \\
Ni    &  7   & -9   & -1  &   -0   &  -1 \\
Cu    &  9   & -10  & -0  &   -0   &  -0 \\
Zn    &  8   & -8   & 0   &   0    &  0  \\
Sr    &  10  & -9   & 2   &   -1   &  -0 \\
Y     &  6   & -6   & 7   &   -6   &  -0 \\
Zr    &  5   & -5   & 7   &   -7   &  -0 \\
Ba    &  9   & -9   & 3   &   -3   &  -1 \\
La    &  8   & -5   & 8   &   -5   &  1  \\
Ce    &  7   & -7   & 6   &   -6   &  0  \\
Nd    &  9   & -8   & 7   &   -6   &  -0 \\
Sm    &  8   & -7   & 6   &   -6   &  -0 \\
Eu    &  8   & -8   & 6   &   -6   &  0  \\
\hline
\end{tabular}

\end{table}

An erroneous \teff leads to an error in abundances $\leq$ 0.1 dex
with almost all elements affected in the same way except C, O, and S
which react in the opposite direction (see Fig.~\ref{error} b and
c). 

It is interesting to note that an erroneous \lgg has almost no
effect on abundances of iron peak elements (Cr to Zn), and leads to an
error $\leq$ 0.1 dex for the other elements. 

When combining both errors, the total error is always $\leq$ 0.15 dex,
as elements that are very sensitive to temperature are not very
sensitive to gravity and conversely, at least for the adopted temperature and
gravity.

\subsection{Errors resulting from wrong continuum adjustment}

Another source of error comes from bad continuum adjustment. 
When rotational velocity increases, it becomes harder to find
continuum points. The use of echelle spectra is helpful because the whole
spectrum from 3900\AA~to 6820\AA~is at our disposal. The normalization
method is presented in Paper I but can be summarized as follows : the
spectrum is divided into 6 parts, with one common order between
parts, and each part is normalized separately. Each part has a length
of more that 400\AA~and was chosen so that there are some continuum
points near the borders. The continuum is fitted with the
\texttt{continuum} function of IRAF whose parameters were defined
using slow rotators, mainly the Sun and Vega. 

Even when working with high S/N ratios, the main problem for continuum
estimation is rotational velocity. It is relatively easy to
normalize a spectrum for a star with a \vsini$\leq$40 \kms, because
lines are not too broad. When rotational velocity increases, lines
become broader and it is hard to find lines that are not
blended. Blends become so important that continuum points vanish. Selecting
wide parts allows to preserve continuum 
points even for large rotational velocities. 

\begin{figure*}[htbp]
  \centering
  \mbox{
        \subfigure[\vsini = 10 ; S/N = 200 ;
	cont~=~1.01]{\includegraphics[width=6.cm]{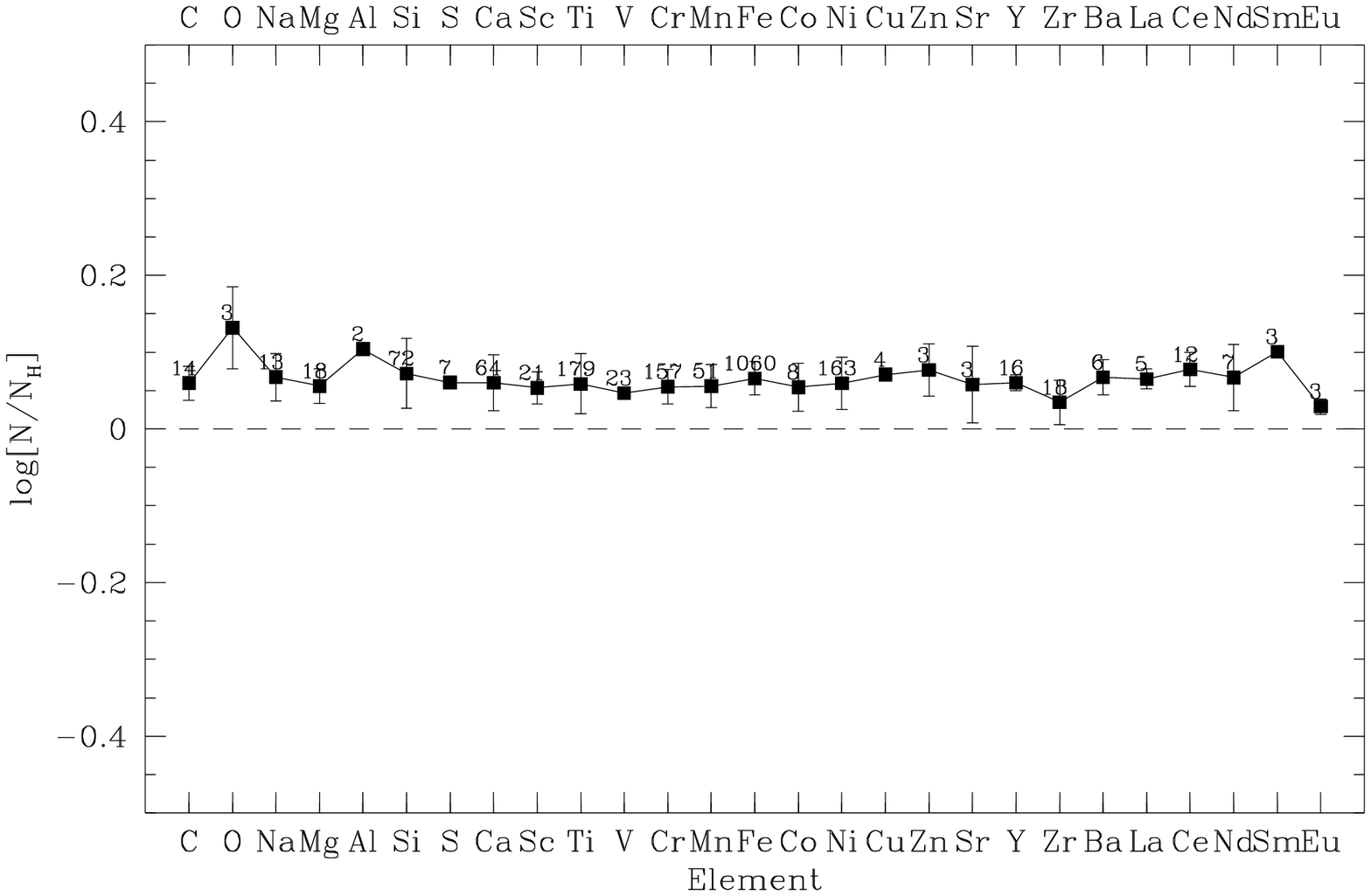}}
        \subfigure[\vsini = 10; S/N = 200 ;
	cont~=~1]{\includegraphics[width=6.cm]{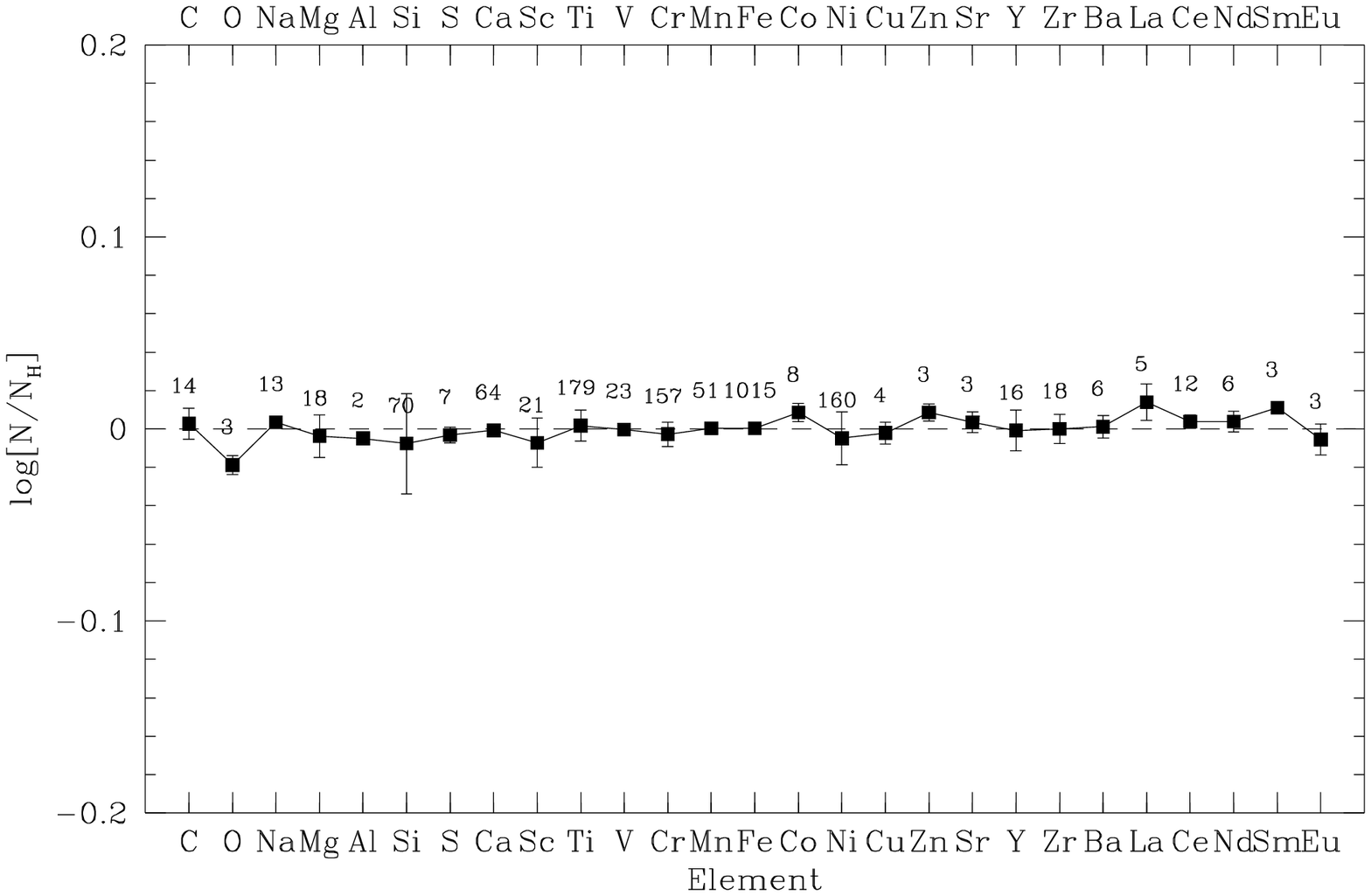}}
        \subfigure[\vsini = 10 ; S/N = 200 ;
	cont~=~0.99]{\includegraphics[width=6.cm]{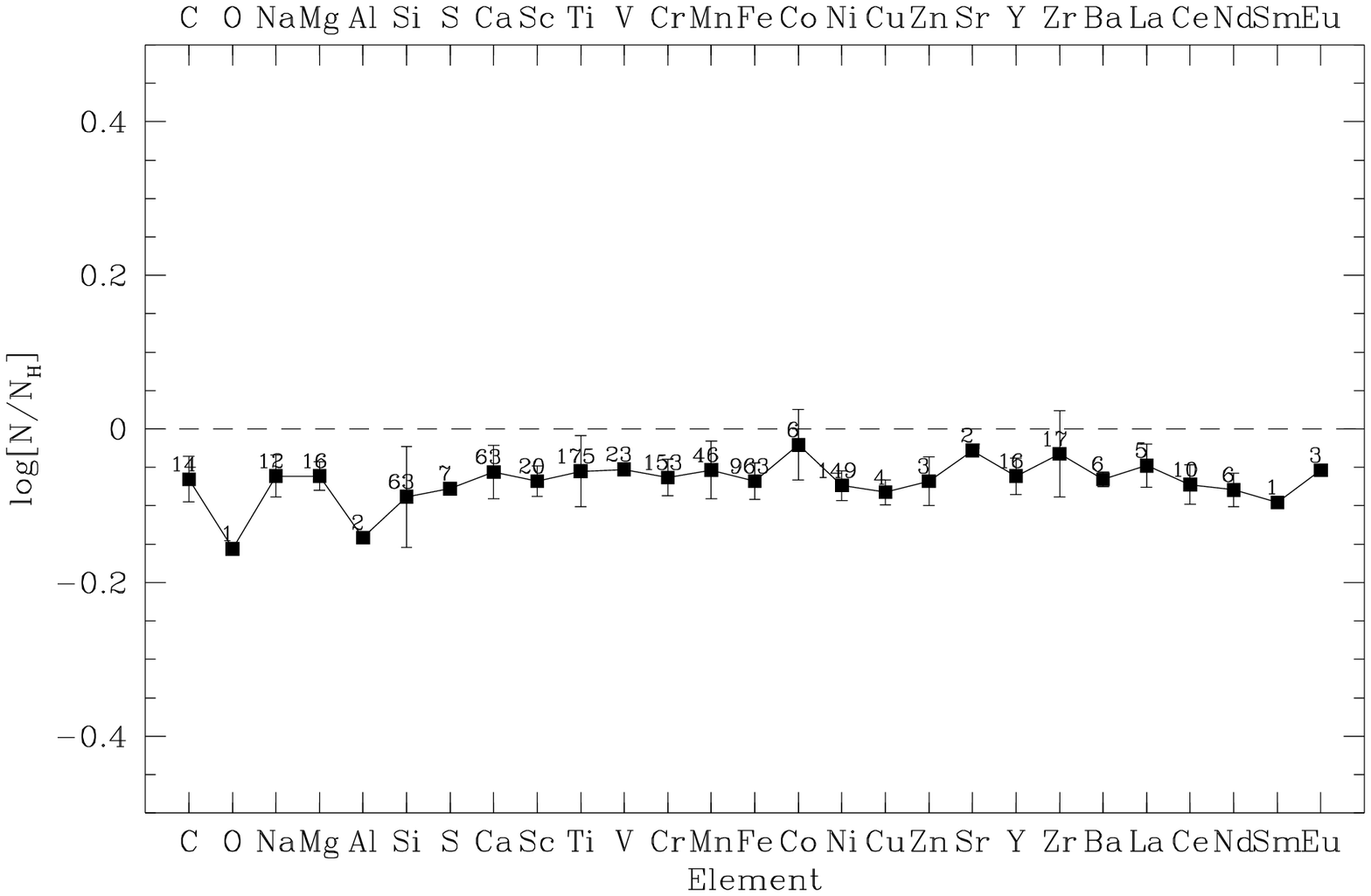}}
  }
  \mbox{
        \subfigure[\vsini = 50 ; S/N = 200 ;
	cont~=~1.01]{\includegraphics[width=6.cm]{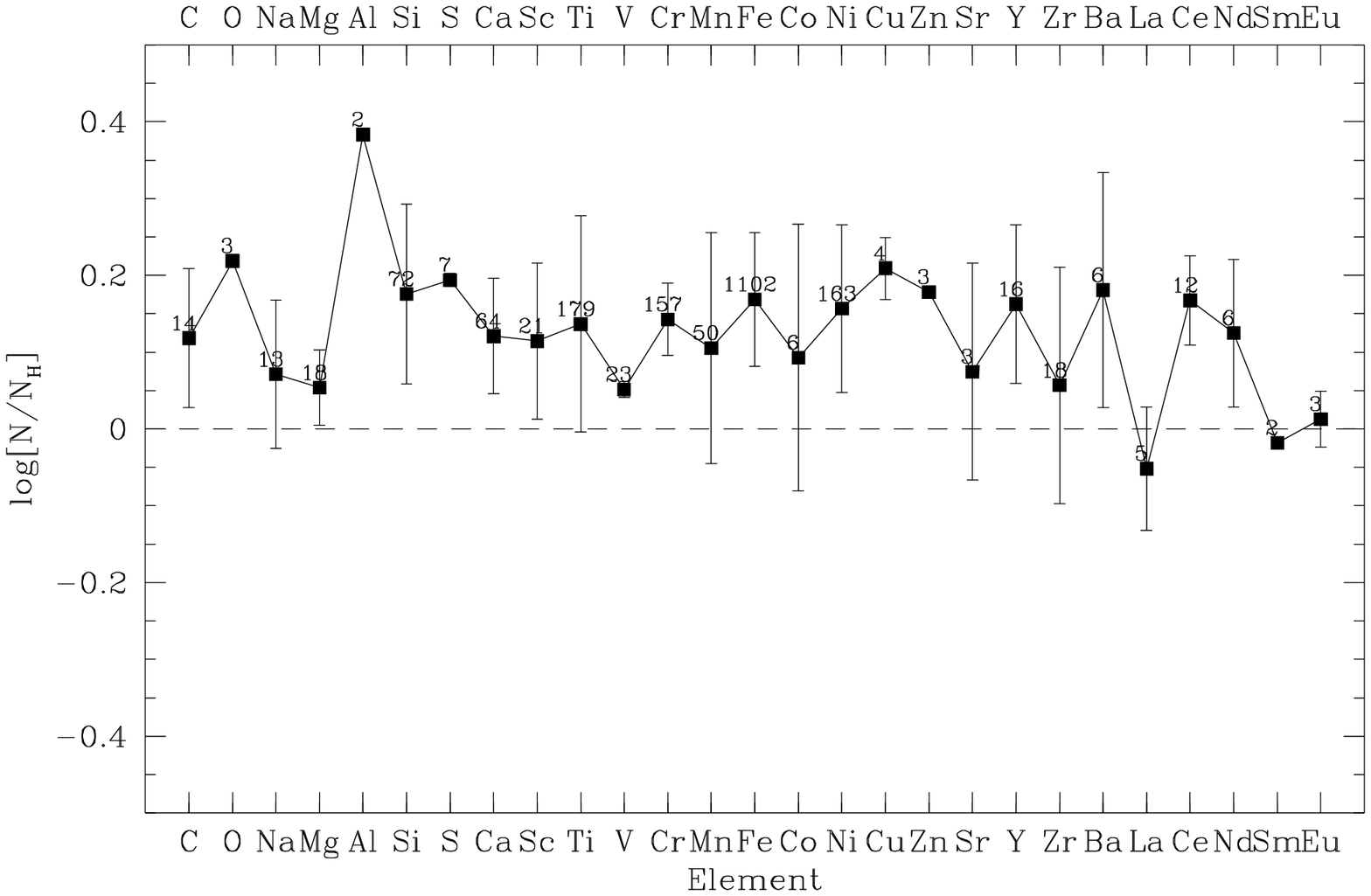}}
        \subfigure[\vsini = 50; S/N = 200 ;
	cont~=~1]{\includegraphics[width=6.cm]{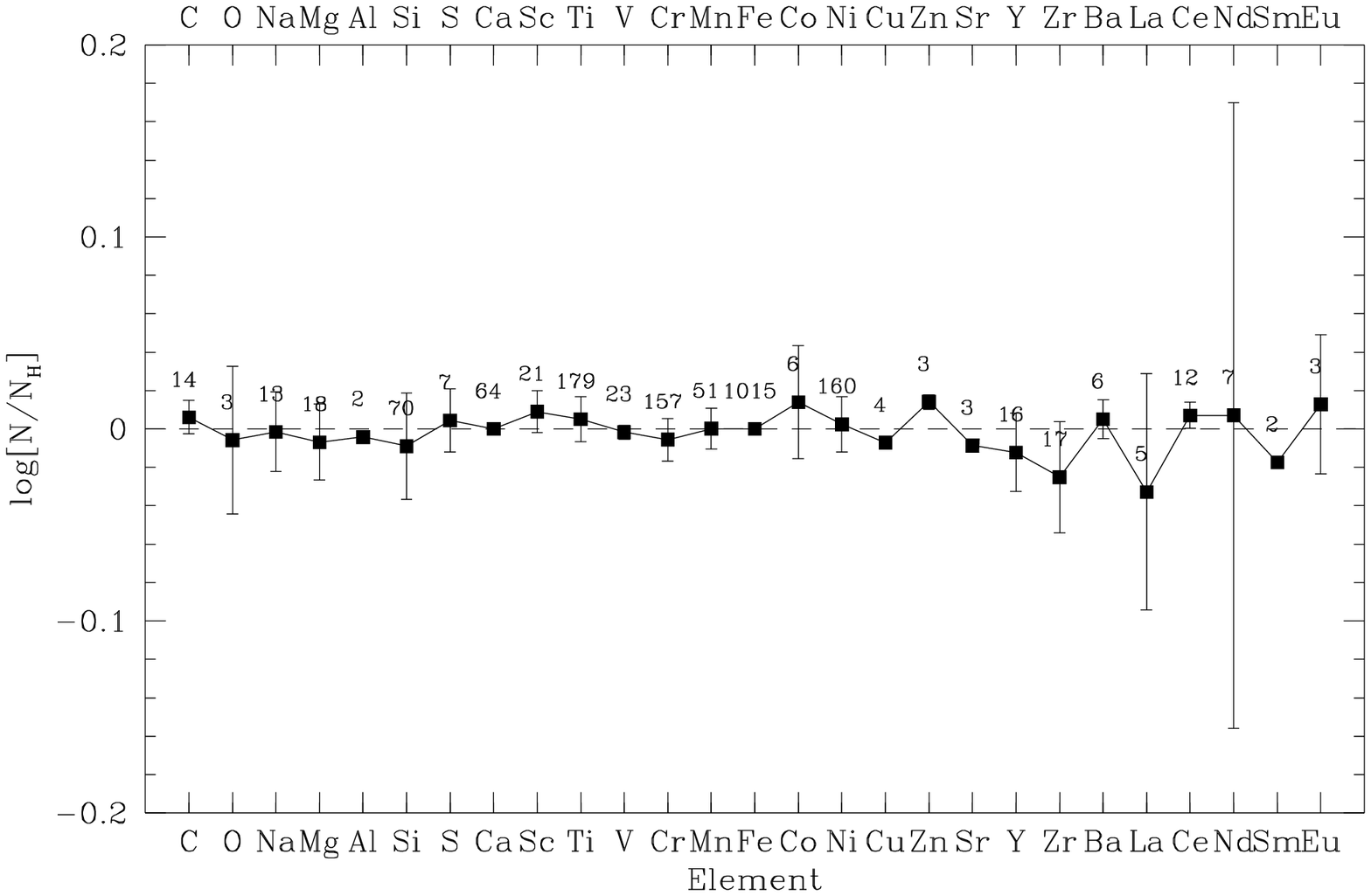}}
        \subfigure[\vsini = 50 ; S/N = 200 ;
	cont~=~0.99]{\includegraphics[width=6.cm]{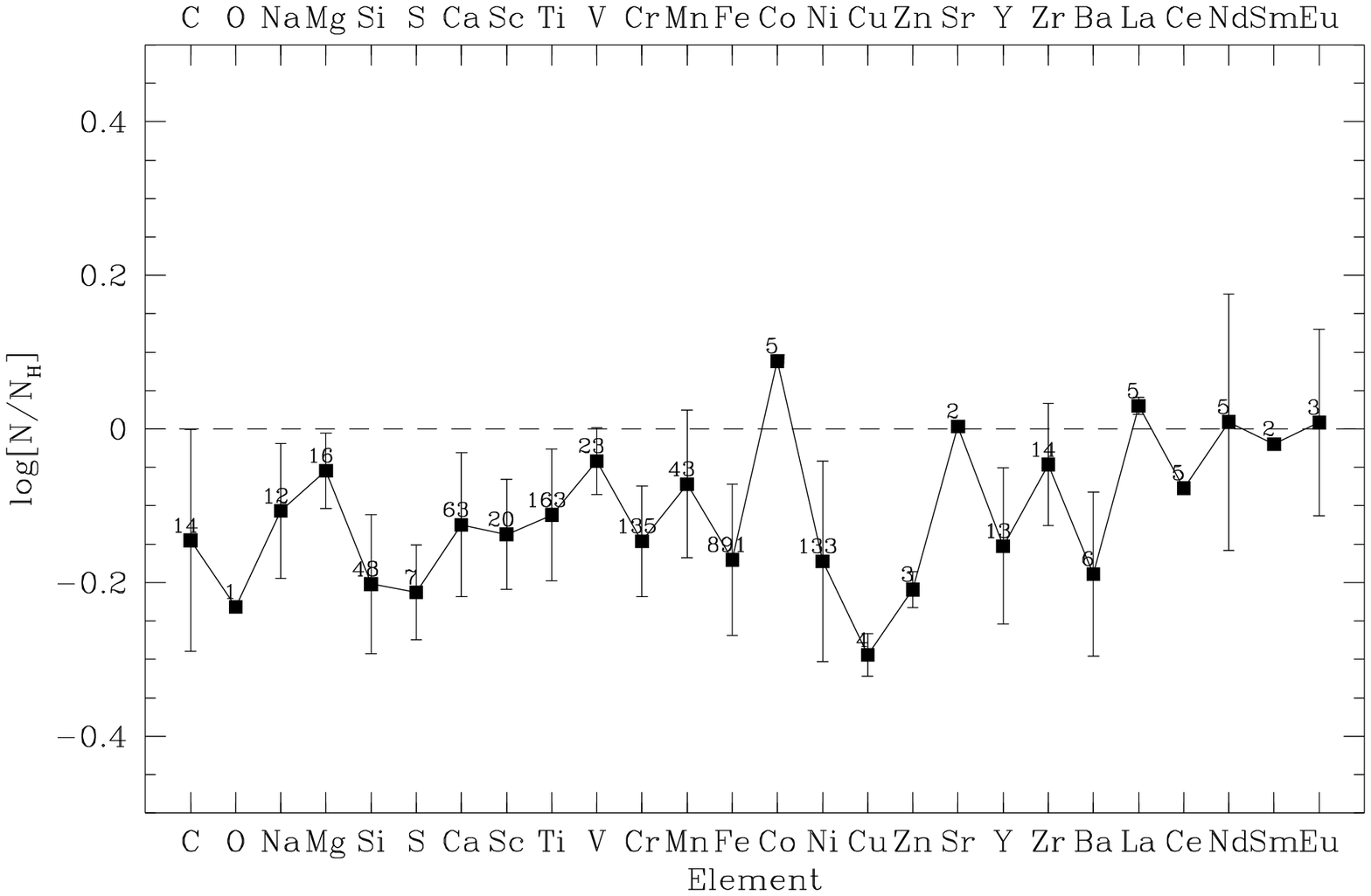}}
  }
  \mbox{
        \subfigure[\vsini = 100 ; S/N = 200 ;
	cont~=~1.01]{\includegraphics[width=6.cm]{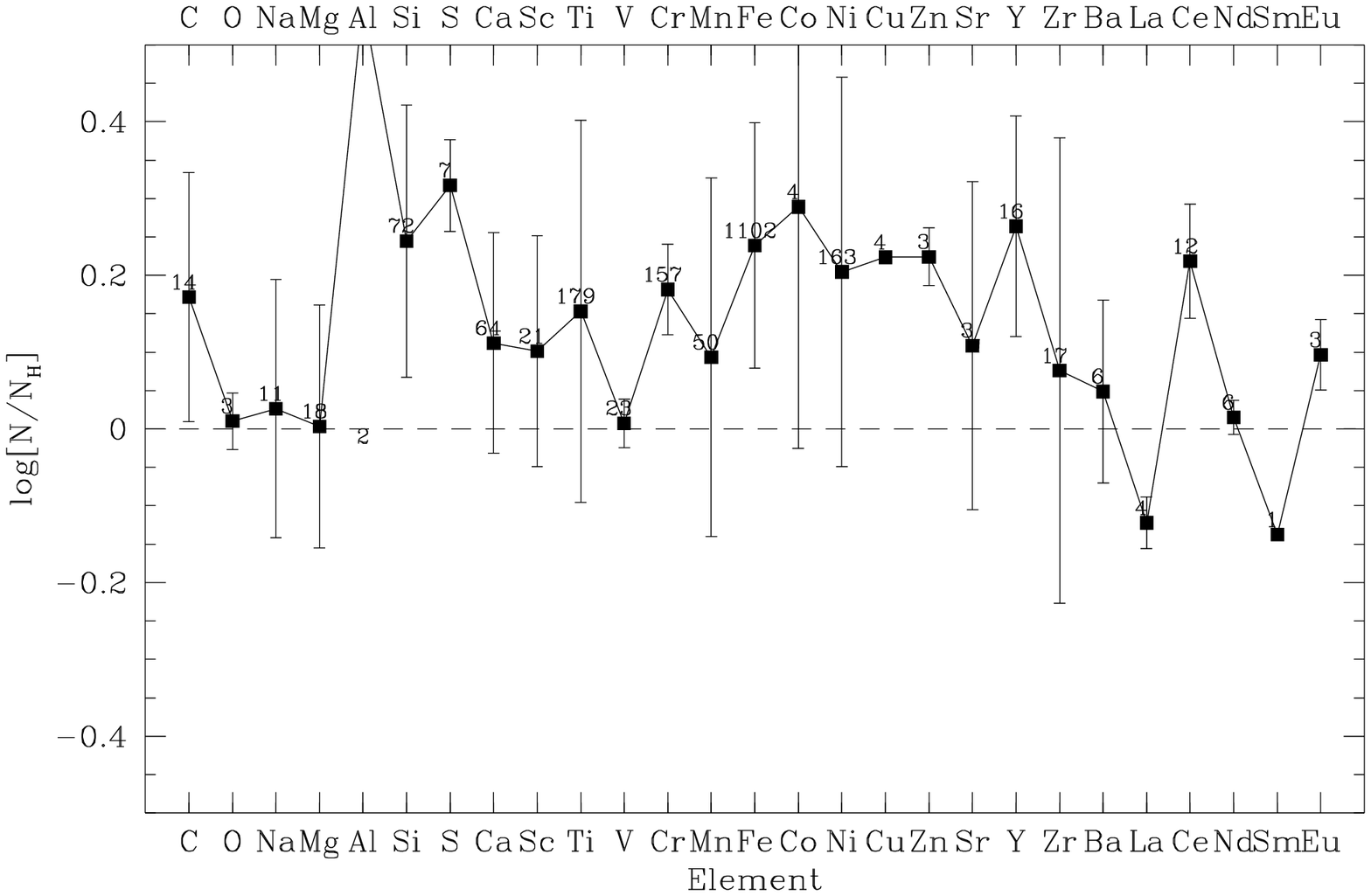}}
        \subfigure[\vsini = 100 ; S/N = 200 ;
	cont~=~1]{\includegraphics[width=6.cm]{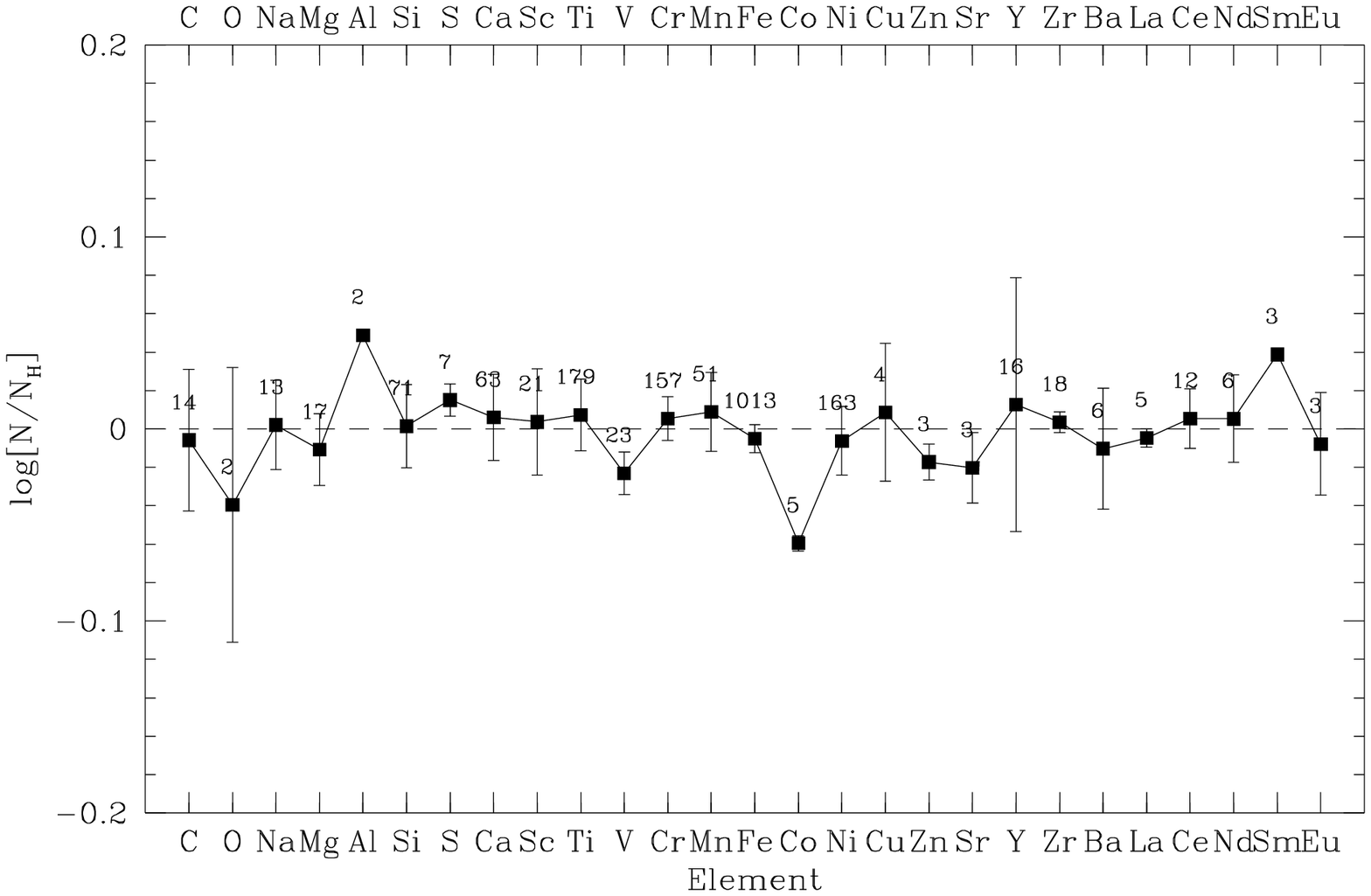}}
        \subfigure[\vsini = 100 ; S/N = 200 ;
	cont~=~0.99]{\includegraphics[width=6.cm]{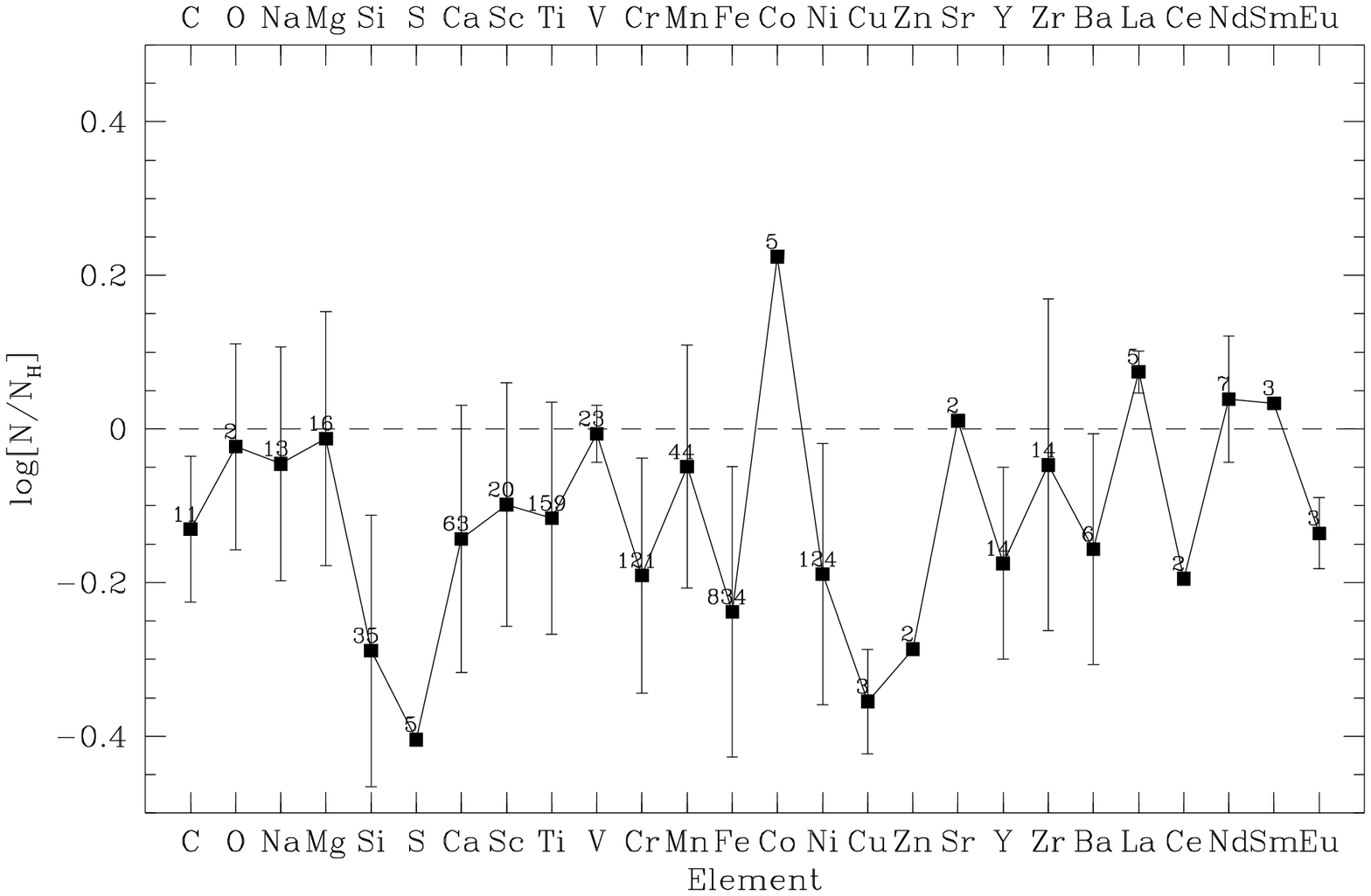}}
  }
  \mbox{
        \subfigure[\vsini = 150 ; S/N = 200 ;
	cont~=~1.01]{\includegraphics[width=6.cm]{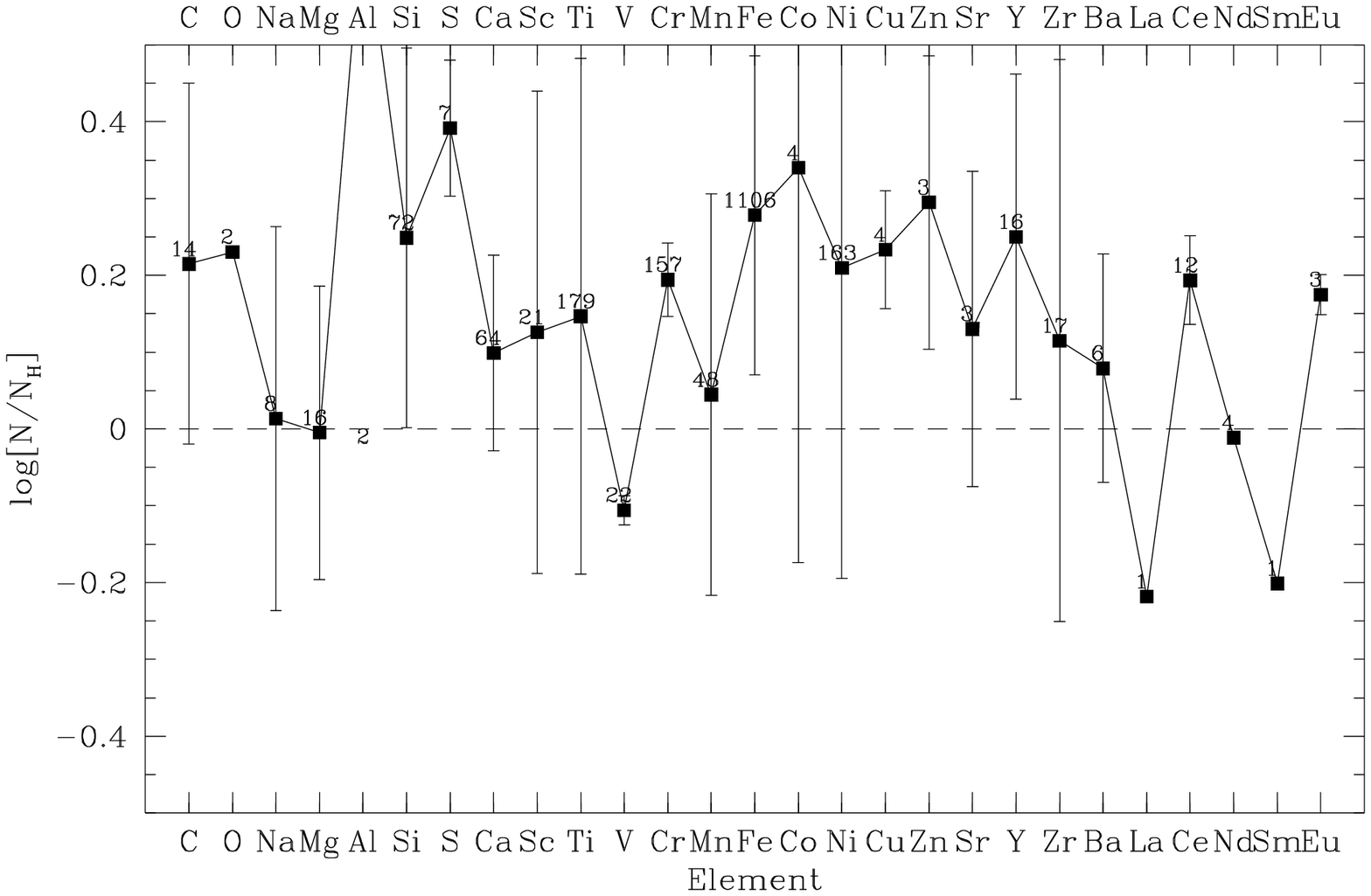}}
        \subfigure[\vsini = 150 ; S/N = 200 ;
	cont~=~1]{\includegraphics[width=6.cm]{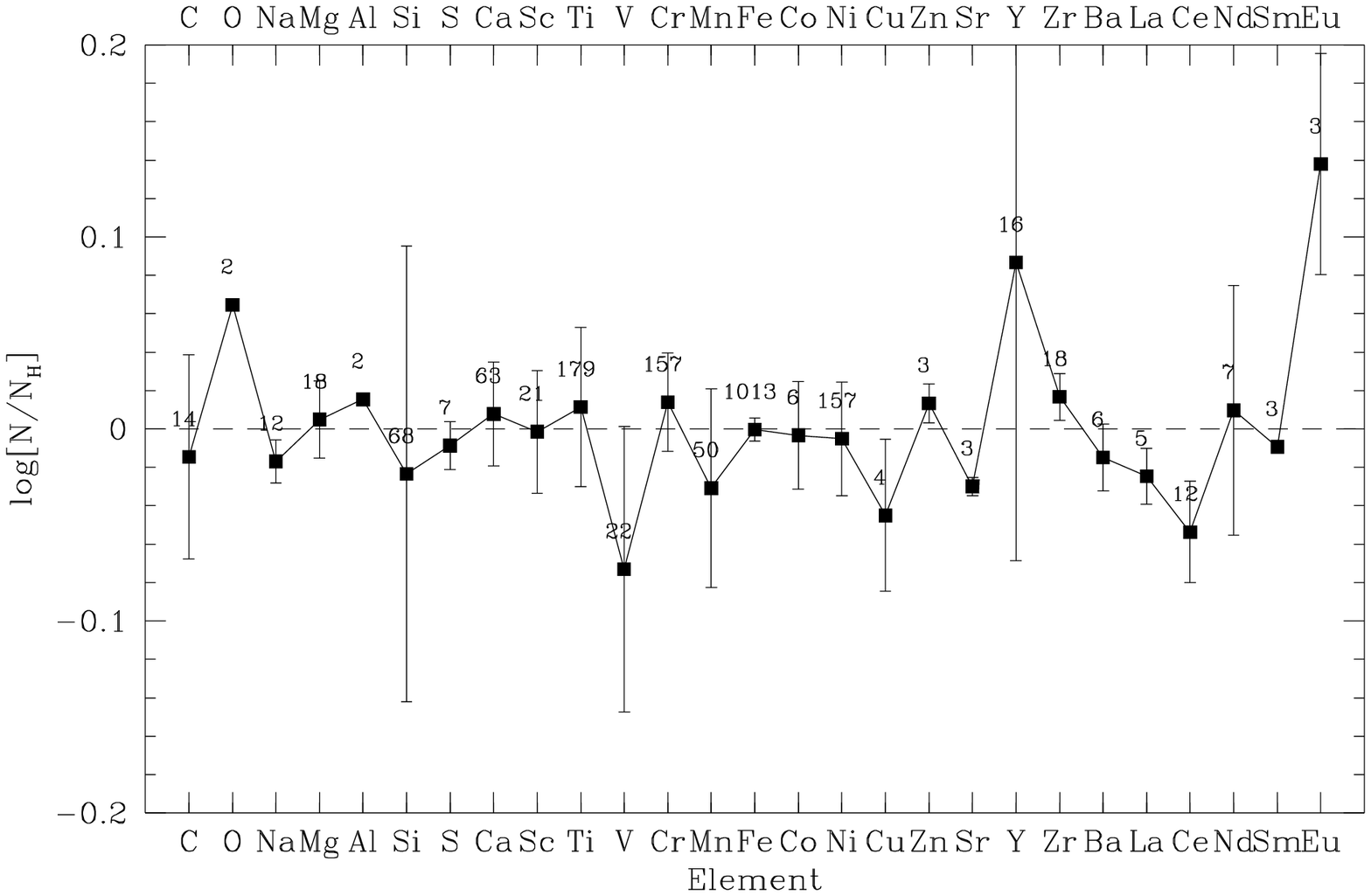}}
        \subfigure[\vsini = 150 ; S/N = 200 ;
	cont~=~0.99]{\includegraphics[width=6.cm]{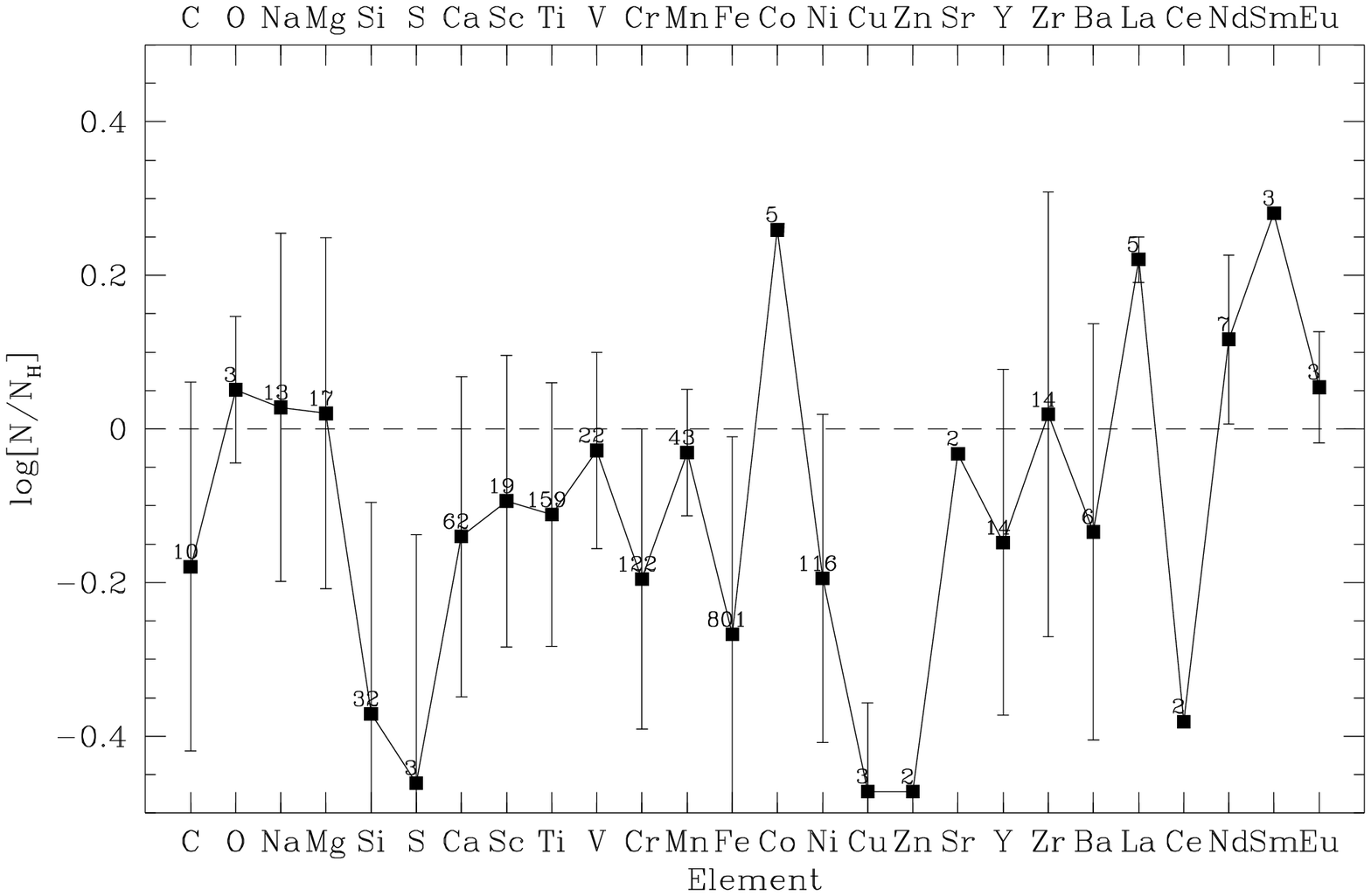}}
  }
  \caption{Abundances with respect to the Sun for different continuum heights. 
    Error bars represent the sigma of the
    average abundance value of the 7 parts of the spectrum (see
    Paper I).}
  \label{contfig}
\end{figure*}

Simulations were done on synthetic spectra having the same parameters
as in previous section, but with \vsini = 10, 50, 100, 150 and 200 \kms.
Moreover, in order to simulate 
as closly as possible real spectra, gaussian noise was added to
obtain a reference 
spectrum with S/N = 200, which corresponds to the minimum observed
value. Finally, the spectra were scaled by a factor 0.99 and 1.01, so that the
continuum level is respectively too high (by a factor 1.01) and too low (by a
factor 0.99). For larger differences, it 
becomes obvious that the continuum is false. Results of these test are
presented in Fig.~\ref{contfig} and in Table~\ref{conttab}.

\begin{table*}
\caption{Difference between adjusted  and reference abundances in
  hundredth of dex for different rotational velocities and different 
  continuum height.}
\label{conttab}
\begin{center}
\begin{tabular}{c||c|c|c||c|c|c||c|c|c||c|c|c|}
        & \multicolumn{3}{c||}{\vsini=10\kms} & \multicolumn{3}{c||}{\vsini=50\kms}
        & \multicolumn{3}{c||}{\vsini=100\kms}& \multicolumn{3}{c|}{\vsini=150\kms}\\

        & 0.99 & 1. & 1.01 &  0.99 & 1. & 1.01 & 0.99 & 1. & 1.01 & 0.99 & 1. & 1.01 \\
\hline
\vrad  &  0   & 0    & 0   &   0   &  0    & 0     &  0    &  0   &  0    &  0    &  0   &  0     \\
\vsini &  0   & 0    & 0   &   0   &  0    & 0     &  0    &  0   &  0    &  0    &  0   &  0     \\
\vmic &   0   & 0    & 0   &   0   &  0    & 0     &  0    &  0   &  0    &  0    &  0   &  0     \\
C     &   6   & 0    & -7  &   12  &  1    & -15   &  17   &  -1  &  -13  &  21   &  -1  &  -18   \\
O     &   13  & -2   & -16 &   22  &  -1   & -23   &  1    &  -4  &  -2   &  23   &  6   &  5     \\
Na    &   7   & 0    & -6  &   7   &  -0   & -11   &  3    &  0   &  -5   &  1    &  -2  &  3     \\
Mg    &   6   & -0   & -6  &   5   &  -1   & -5    &  0    &  -1  &  -1   &  -0   &  0   &  2     \\
Al    &   10  & -0   & -14 &   38  &  -0   &       &  56   &  5   &       &  68   &  2   &        \\
Si    &   7   & -1   & -9  &   18  &  -1   & -20   &  24   &  0   &  -29  &  25   &  -2  &  -37   \\
S     &   6   & -0   & -8  &   19  &  0    & -21   &  32   &  2   &  -40  &  39   &  -1  &  -46   \\
Ca    &   6   & -0   & -6  &   12  &  -0   & -12   &  11   &  1   &  -14  &  10   &  1   &  -14   \\
Sc    &   5   & -1   & -7  &   11  &  1    & -14   &  10   &  0   &  -10  &  13   &  -0  &  -9    \\
Ti    &   6   & 0    & -6  &   14  &  1    & -11   &  15   &  1   &  -12  &  15   &  1   &  -11   \\
V     &   5   & -0   & -5  &   5   &  -0   & -4    &  1    &  -2  &  -1   &  -11  &  -7  &  -3    \\
Cr    &   6   & -0   & -6  &   14  &  -1   & -15   &  18   &  1   &  -19  &  19   &  1   &  -20   \\
Mn    &   6   & 0    & -5  &   11  &  0    & -7    &  9    &  1   &  -5   &  4    &  -3  &  -3    \\
Fe    &   7   & 0    & -7  &   17  &  -0   & -17   &  24   &  -1  &  -24  &  28   &  -0  &  -27   \\
Co    &   5   & 1    & -2  &   9   &  1    & 9     &  29   &  -6  &  22   &  34   &  -0  &  26    \\
Ni    &   6   & -0   & -7  &   16  &  0    & -17   &  20   &  -1  &  -19  &  21   &  -1  &  -19   \\
Cu    &   7   & -0   & -8  &   21  &  -1   & -29   &  22   &  1   &  -35  &  23   &  -5  &  -47   \\
Zn    &   8   & 1    & -7  &   18  &  1    & -21   &  22   &  -2  &  -29  &  29   &  1   &  -47   \\
Sr    &   6   & 0    & -3  &   7   &  -1   & 0     &  11   &  -2  &  1    &  13   &  -3  &  -3    \\
Y     &   6   & -0   & -6  &   16  &  -1   & -15   &  26   &  1   &  -17  &  25   &  9   &  -15   \\
Zr    &   3   & -0   & -3  &   6   &  -3   & -5    &  8    &  0   &  -5   &  11   &  2   &  2     \\
Ba    &   7   & 0    & -7  &   18  &  1    & -19   &  5    &  -1  &  -16  &  8    &  -1  &  -13   \\
La    &   7   & 1    & -5  &   -5  &  -3   & 3     &  -12  &  -0  &  7    &  -22  &  -2  &  22    \\
Ce    &   8   & 0    & -7  &   17  &  1    & -8    &  22   &  1   &  -20  &  19   &  -5  &  -38   \\
Nd    &   7   & 0    & -8  &   12  &  1    & 1     &  1    &  1   &  4    &  -1   &  1   &  12    \\
Sm    &   10  & 1    & -10 &   -2  &  -2   & -2    &  -14  &  4   &  3    &  -20  &  -1  &  28    \\
Eu    &   3   & -1   & -5  &   1   &  1    & 1     &  10   &  -1  &  -14  &  17   &  14  &  5     \\
\hline
\end{tabular}
\end{center}
\end{table*}

It is interesting to note that for the correct continuum definition,
almost all adjusted abundances are well within 0.1 dex of their input
values, even for high
\vsini (see central column of Fig.~\ref{contfig}). Moreover the
adjustment with the correct value allows to select ``sensitive"
elements, such as Al, Si, S, V, Cu, Y, and Nd. These elements have very
few and very weak lines, generally blended with much stronger lines. 

Another interesting point is the shape of the abundance pattern
obtained with a scaled continuum. This shape remains very constant when the
rotational velocity increases, but its amplitude increases with this
velocity. 

Finally, this simulation can be interpreted as yielding the maximum possible
error, because all adjusted parts were scaled by one and the same factor. In
reality, even for \vsini = 150 \kms, only the first two parts
($\lambda < 4900$\AA) can be
affected by the lack of continuum points, the 5 remaining parts being
much easier to normalize. Therefore it is possible to conclude that
the influence of the normalization on real spectra will be well below
the values listed in Table~\ref{conttab}.

\section{Results}\label{sec:res}
All results are listed in Tables 5 and 6 available at the CDS. The first table
lists, for each star, the radial velocity, the microturbulence and the projected
rotational velocity. The second one lists the abundances. In 12 cases, there
are two spectra per star -- hence two lines per star in Tables 5 and 6 -- either
because the same object was observed once with ELODIE and once with CORALIE (6
cases), or because the star was observed twice with a given instrument (6 more
cases).

\subsection{Abundances}
As the sample covers a large temperature domain, it is difficult to
summarize the results. For early A-type stars, only a few elements
show lines in the spectrum while for F-type stars, it is possible to
derive abundances for 28 elements. Table~6, available at CDS, contains 
the detailed abundances for each star and the number of lines used
for the determination.

\begin{figure*}[htbp]
   \mbox{
        \subfigure{\includegraphics[width=6.cm]{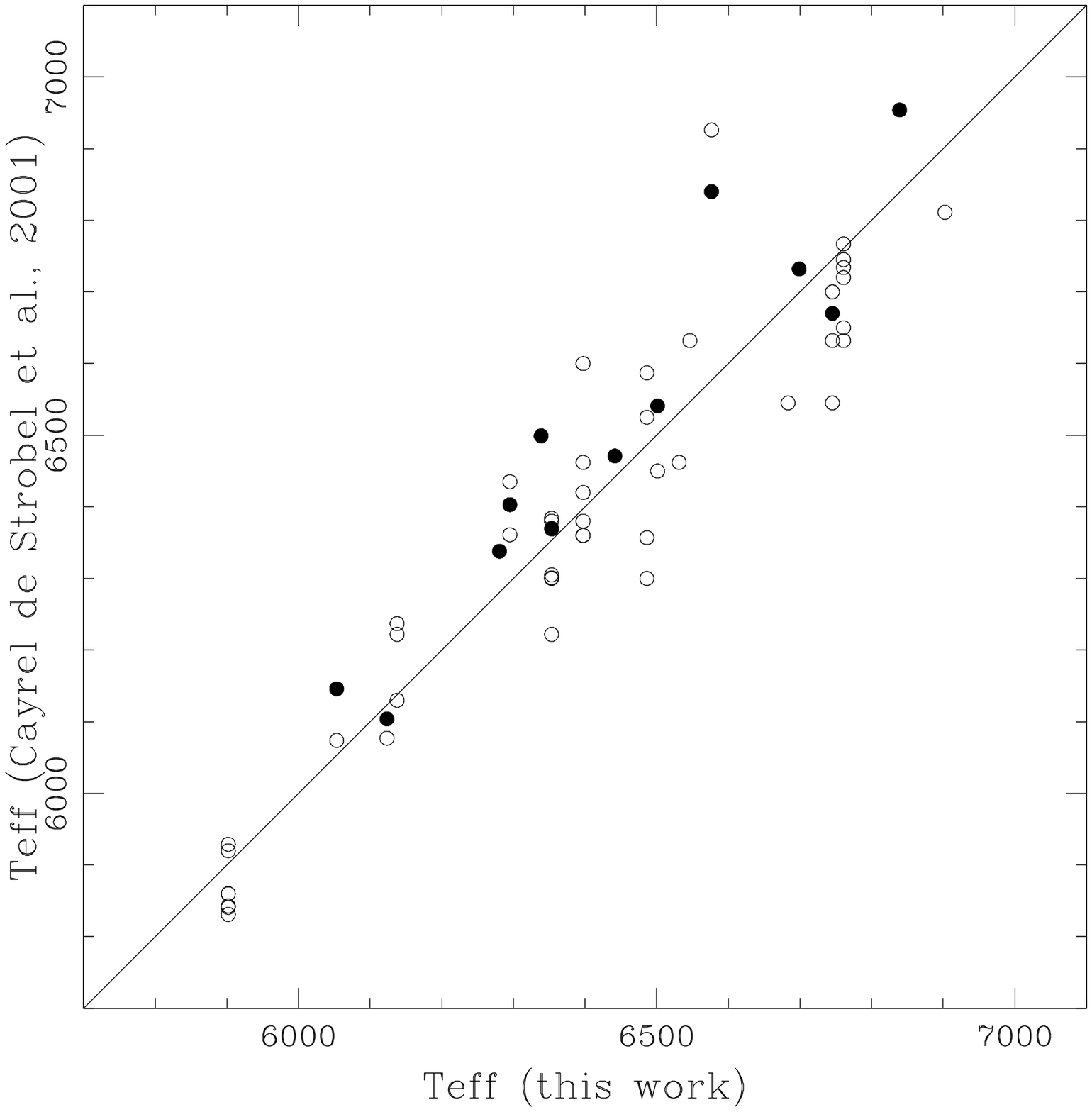}}
        \subfigure{\includegraphics[width=6.cm]{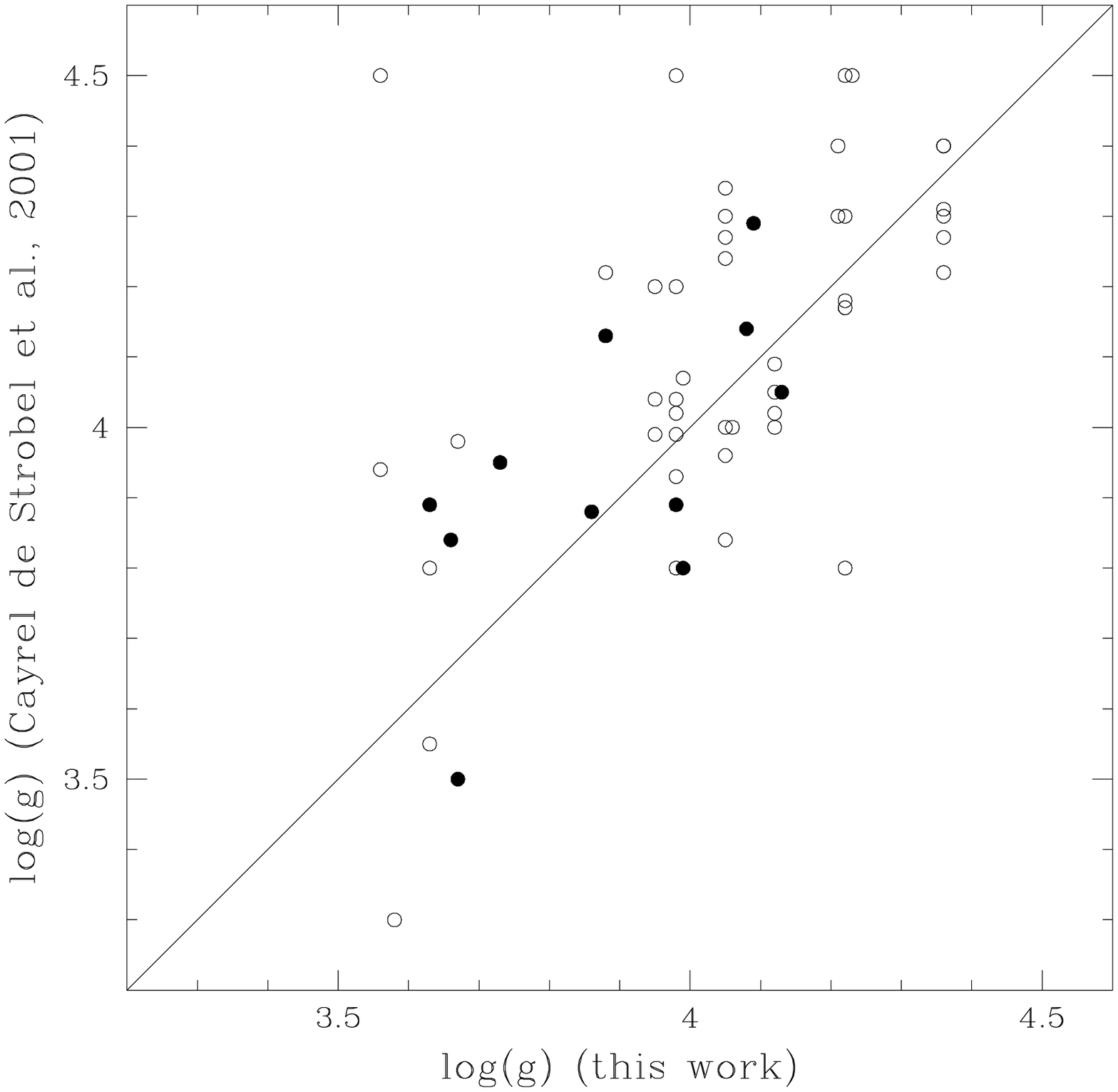}}
        \subfigure{\includegraphics[width=6.cm]{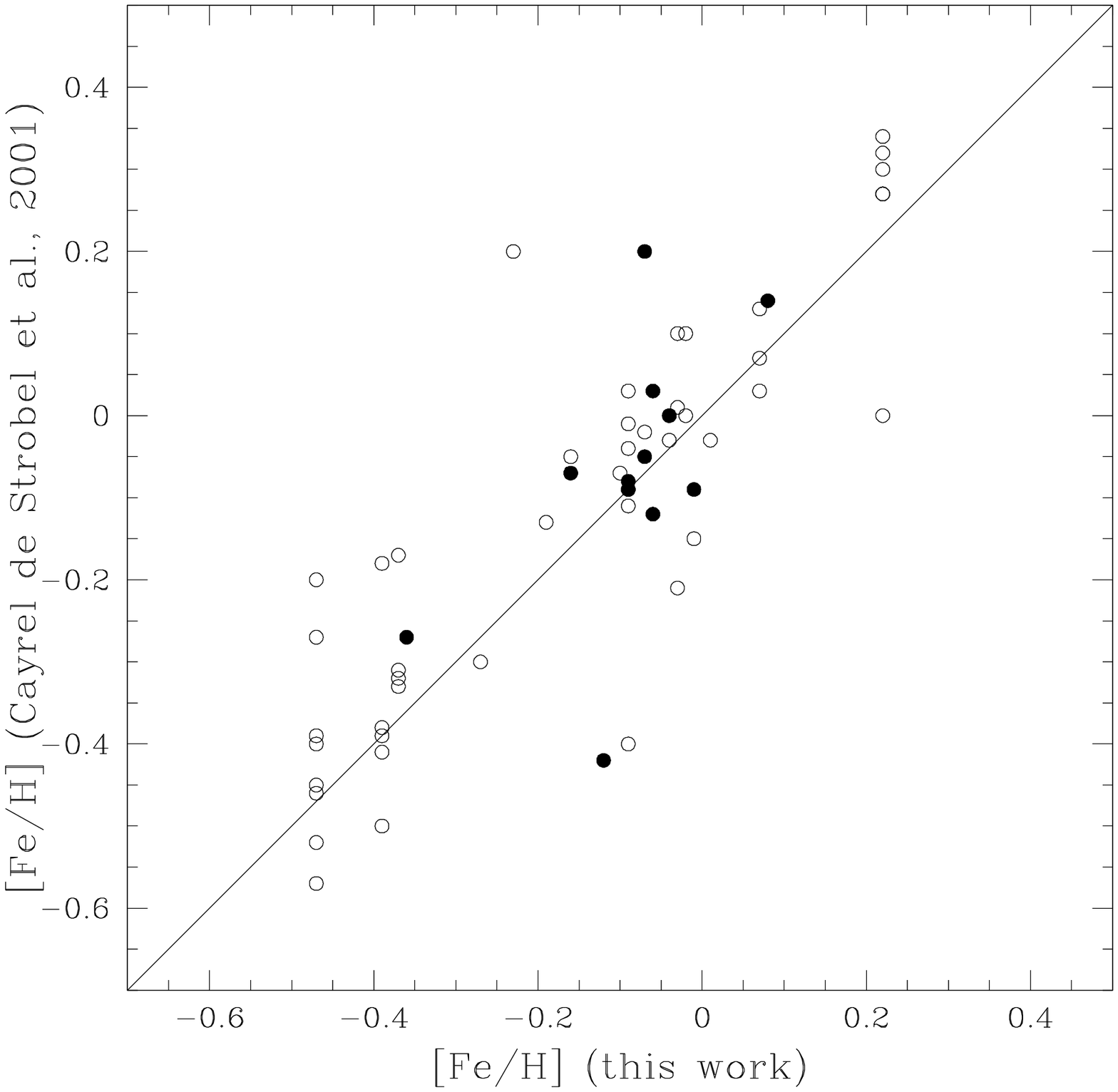}}
 }
  \caption{Comparison between \teff, \lgg, $\left[\frac{Fe}{H}\right]$
    from this work and from the $\left[\frac{Fe}{H}\right]$ catalogue of
    Cayrel de Strobel et al. (\cite{cayrel}). Filled circles distinguish 
    Balachandran's (\cite{balach}) contribution, open circles representing
    determinations by other authors.The straight line is not an adjustment but
    represents the equality of both values}
  \label{caycomp}
\end{figure*}

24 stars of the sample are common with the catalogue of
$\left[\frac{Fe}{H}\right]$ done by Cayrel de Strobel et
al. (\cite{cayrel}). Comparisons for \teff, \lgg and
$\left[\frac{Fe}{H}\right]$ are illustrated in
Fig.~\ref{caycomp}. The agreement is very good. It is intersting to 
notice that 12 stars come from a single, homogeneous source, Balachandran
(\cite{balach}). This sample alone shows a slight
shift in \teff of less than 100 K and \lgg values which are in good
agreement with ours. $\left[\frac{Fe}{H}\right]$ values are
in good agreement, except for two cases where the difference amounts to about
0.3~dex.

\subsection{Microturbulence}
Microturbulence is an {\it ad hoc} parameter which relates to turbulent motions
on scales smaller than the mean free path of a photon. As commonly done in
practice, we assumed it is constant with optical depth even though it should be
allowed to vary in principle. Our estimates are based on the the four bluer
spectral segments only, since there are too few appropriate lines in the three
redder segments to constrain the adjustment; the microturbulence value
is an average of the four individual 
fitted values. Because of the large
size and homogeneity of our sample, the relation between microturbulence,
effective temperature and luminosity is worth examining.
 
\begin{figure}[htbp]
\resizebox{\hsize}{!}{\includegraphics{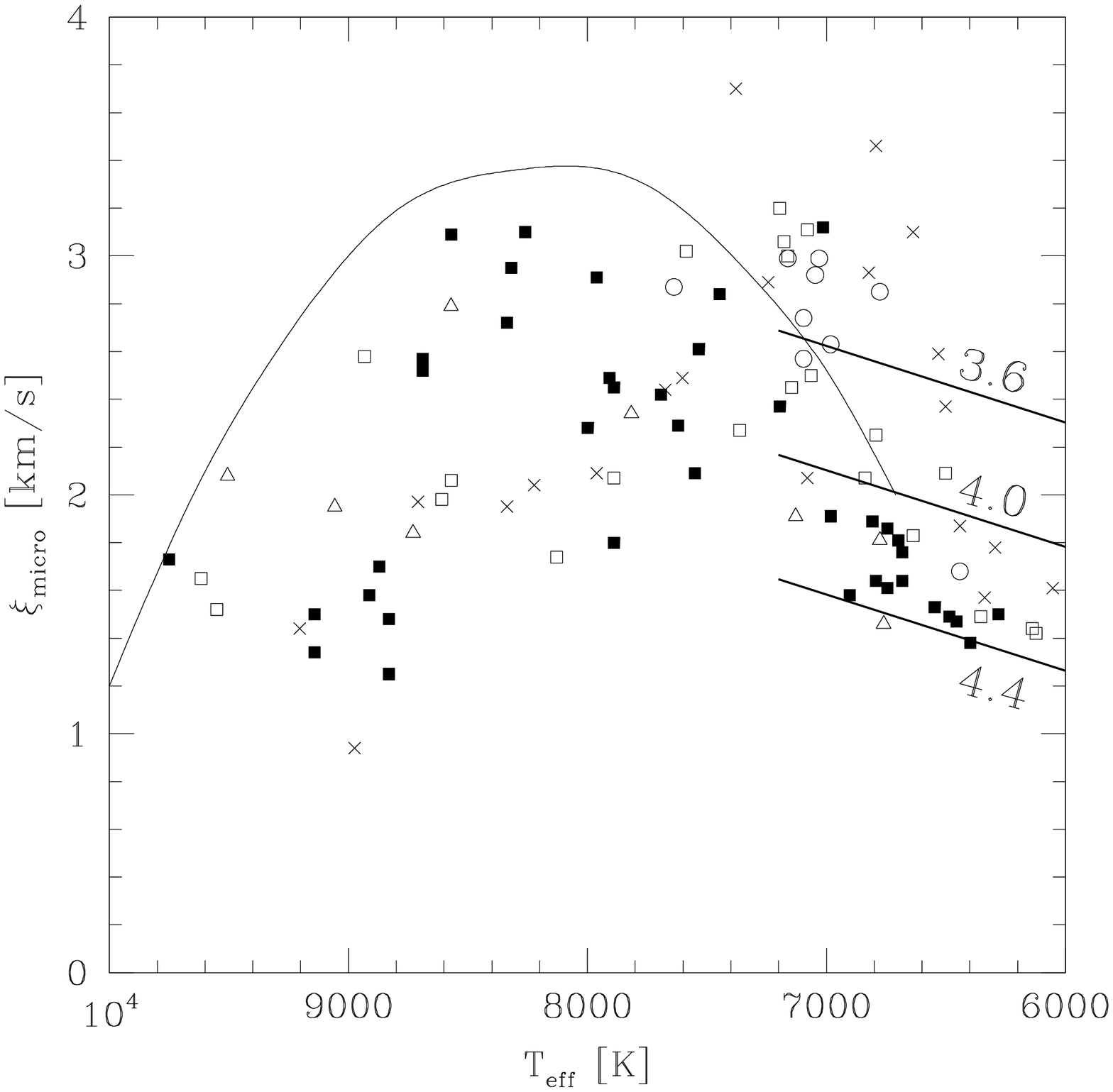}}
\caption{Relation between \teff and \vmic. $\triangle$ for stars with a reliable
\vmic determination (see text): stars with
  \lgg $>$ 4.25. $\blacksquare$ : stars with $4.0<$\lgg$\leq 4.25$.
$\square$ : stars with $3.75<$\lgg$\leq 4.0$.
$\times$ : stars with $3.5<$\lgg$\leq 3.75$.
$\bigcirc$ : stars with \lgg $\leq 3.5$.
The continuous line is the relation of Coupry \& Burkhart (\cite{coupry92}),
while the thick straight lines are Nissen's (\cite{N81}) relation for the three
indicated \lgg values.}
\label{vturb}
\end{figure}

The values are reported in Table~5 available at CDS.
Fig.~\ref{vturb} shows the relation between \teff and \vmic for reliable \vmic
estimates, defined as those for which the standard deviation of the 4 individual
fitted values is smaller than 0.5 \kms.
This relation is compared with the average, hand-drawn one presented in
Coupry \& Burkhart (\cite{coupry92}), which is the continuous curve in
Fig.~\ref{vturb}. The agreement is good for cool stars but, beyond
\teff$=7400$~K, the relation of Coupry \& Burkhart becomes an upper limit to
our values.

The relation of Nissen (\cite{N81}), which is linear in both \teff
and \lgg, is represented by the three straight lines in its domain of validity.
The agreement seems not too bad for low luminosity stars, but our data seem to
indicate that all stars share the same \vmic at low temperature, whatever their
\lgg, while evolved stars separate from the unevolved ones very rapidly as \teff
increases. The \teff dependence of \vmic is much steeper in our case than in
Nissen's, and the slope seems to increase with decreasing \lgg. Qualitatively,
however, we confirm that evolved stars have higher \vmic in this range of \teff.
This is also in qualitative agreement with the results of
Gray et al. (\cite{gray}) for stars cooler than A8.

Strangely enough, the trend reverses for \teff between 8000 and 8800~K, evolved
stars having significantly lower \vmic on average than unevolved ones.
The latter trend is largely due to
the group of stars with \teff larger than 8000 K and \vmic larger than
2.4, which correspond to stars with mass of 1.8 to 2.0 \msol that present iron
and nickel overabundances (see Sect.~\ref{abund} for details). These stars are
very close to the Zero Age Main Sequence (ZAMS) in the HR diagram.
This behaviour was already noted by Gray et al. (\cite{gray}).
Whether \vmic depends on $[M/H]$ remains an open question, and more high quality
data would be needed to better understand what is happening in this diagram.
Let us notice that the observed overabundance of iron cannot be the result of
a bias in the estimate of \vmic, since only an {\it underestimate} of it would 
lead to an overestimate of iron abundance; what we see is, on the contrary,
rather high \vmic values for these metal rich stars.

It is difficult to draw any conclusion about a possible relation between \vmic
and \lgg for the stars hotter than 8800~K because their number is too small. In
addition, the \vmic estimates are less reliable than for cool stars because of
the much smaller number of lines.

\section{Discussion}\label{sec:disc}

\subsection{Rotational velocities}

\subsubsection{Comparison with other works}
\begin{figure*}[htbp]
   \mbox{
        \subfigure{\includegraphics[width=6.cm]{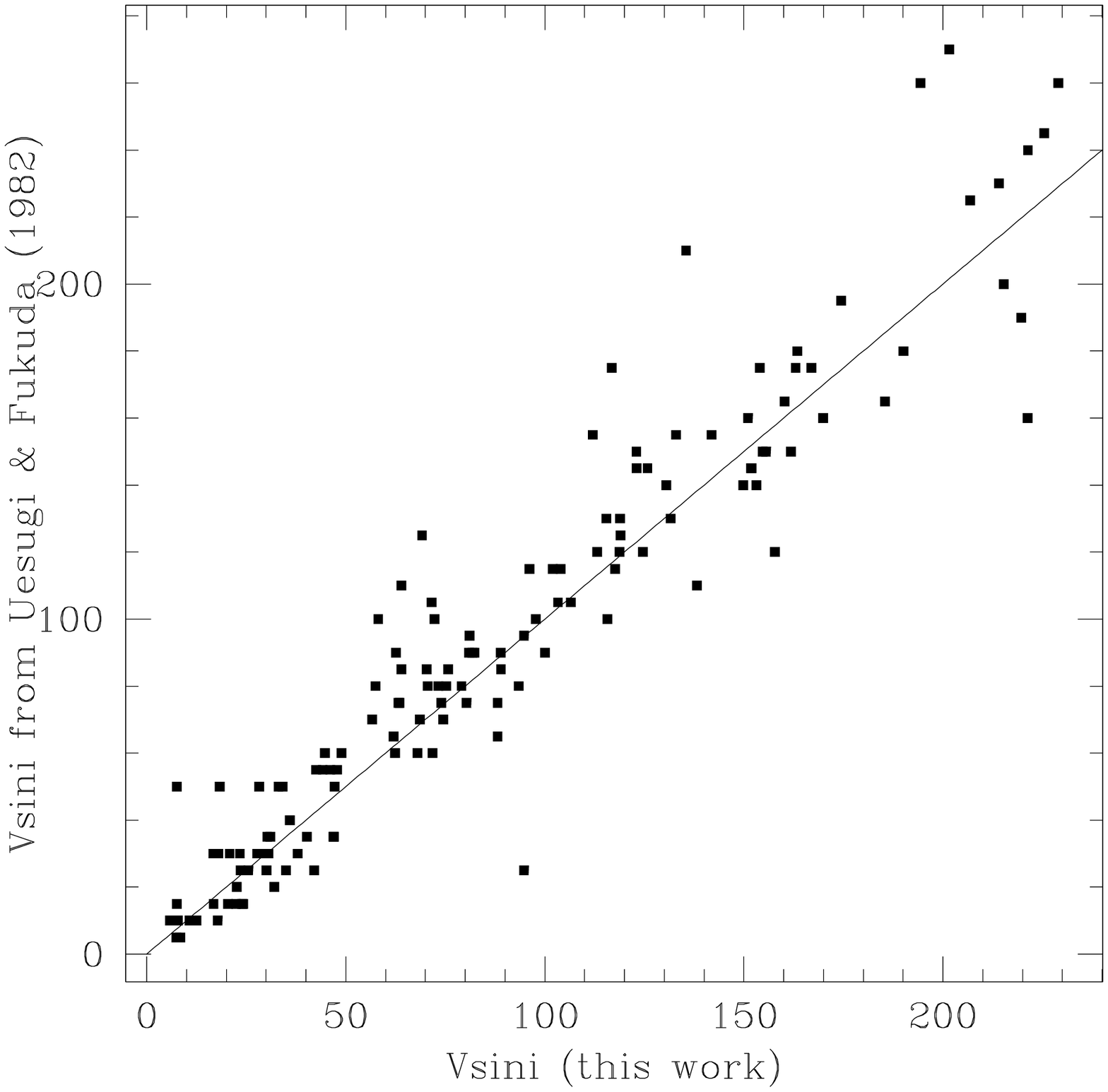}}
        \subfigure{\includegraphics[width=6.cm]{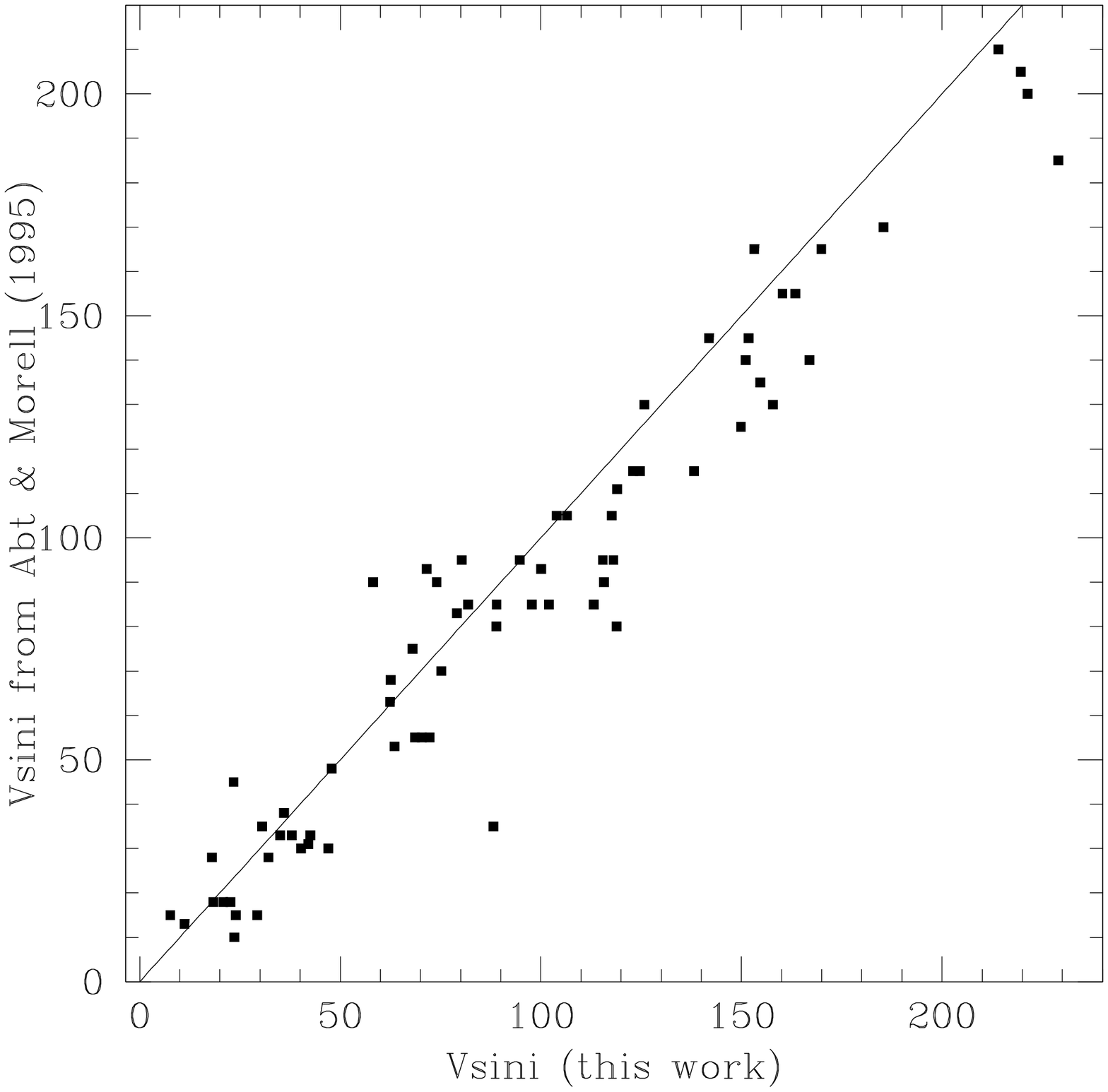}}
        \subfigure{\includegraphics[width=6.cm]{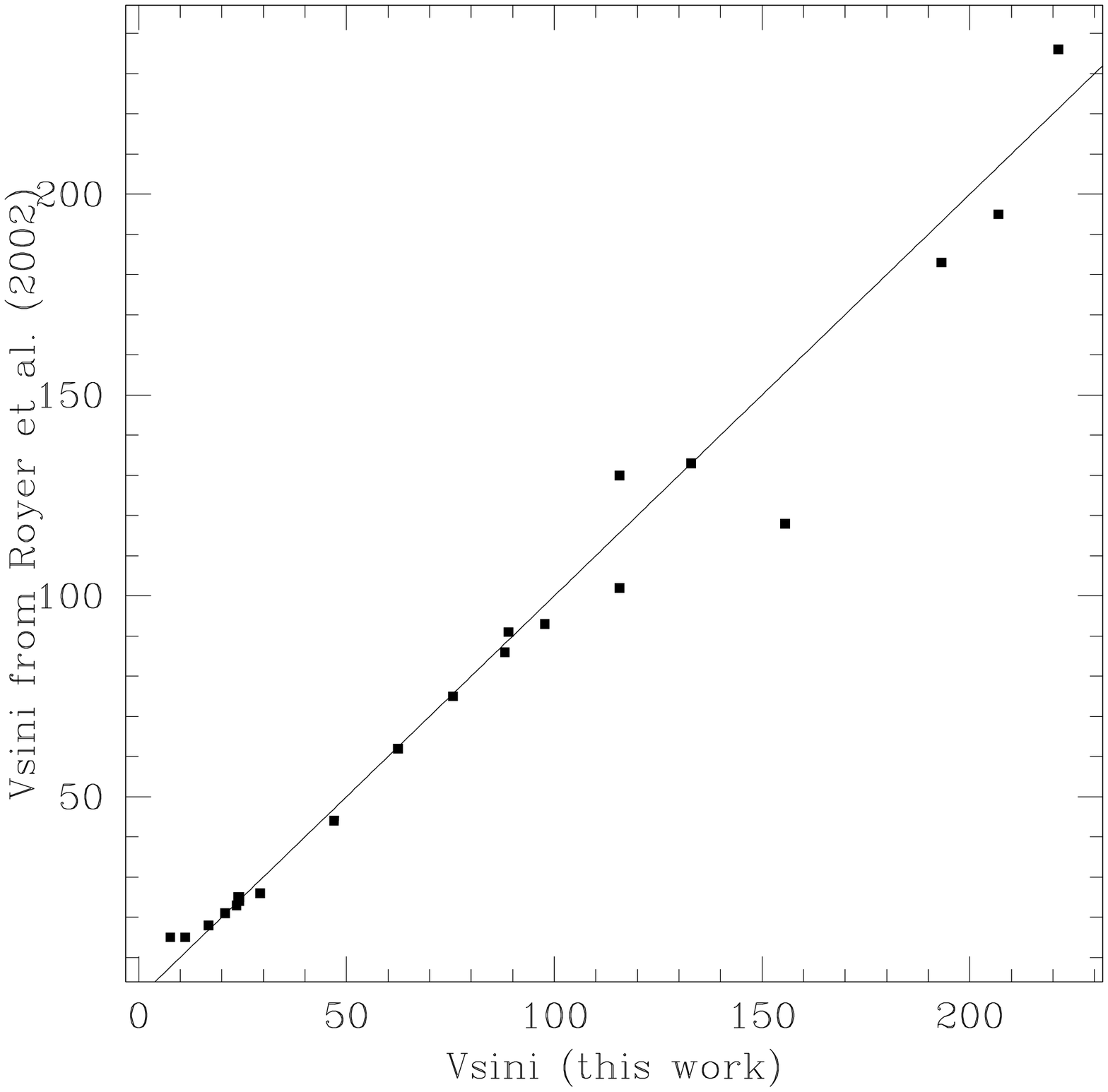}}
 }
  \caption{Comparison between \vsini values from this work and Uesugi
    \& Fukuda (\cite{uesugi}), Abt \& Morell (\cite{abt}) and Royer et 
    al. (\cite{royer}). The straight line is not an adjustment but
    represents the equality of both values.}
  \label{vsinicomp}
\end{figure*}

Fig.~\ref{vsinicomp} illustrates the comparison of our determinations
of \vsini with
works of Uesugi \& Fukuda (\cite{uesugi}), Abt \& Morell
(\cite{abt}), and Royer et al. (\cite{royer}). Our values are slightly larger
than those of Abt \& Morell, especially
for \vsini larger than 100 \kms. However, the trend is in the other
direction when comparing with Uesugi \& Fukuda. Finally, comparison
with Royer et al. is excellent if we neglect one point
($\sim$150 \kms for  
our estimate). This point corresponds to HD 12311. This star
  is still
classified as a dwarf in Simbad even though it has the atmospheric parameters of a
giant. This was already noted by Hauck (\cite{hauck}) using
\textsc{Geneva} photometry,
  and Gray \& Garrison (\cite{gray89}) give a spectral type of F0
  III-IVn. Royer et
al. (\cite{royer}) use Fourier transforms of several line profiles to
estimate \vsini. This technique is very efficient for stars with
well separated lines. In cooler stars, it becomes harder to find
isolated lines to analyze. HD 12311 is precisely one of the coolest stars of
Royer et al. (\cite{royer}). In this case, all lines of their list except one
were unusable. Therefore the difference can be explained by a poor
precision of their determination for this star. 
In this comparison, there are
less points because Royer et al. focus on A type stars.

We conclude that our determination of \vsini are probably even better than those
of Royer et al., because we make use of all lines present in the spectra,
whether they are blended or not. On the other hand, we neglect macroturbulence,
so our \vsini values are slightly biased towards large values in very
slow rotators (\vsini $< 8-10$ \kms).

\subsubsection{Systematic abundance error due to rotation} 

When dealing with large rotational velocities, one can easily
underestimate the continuum level. That leads to underestimated
abundances because lines seem shallower. 

The apparent abundances of three elements are strongly correlated
with \vsini. These elements are Sr, 
Sc, and Na. This is illustrated in Fig.~\ref{vs}, where iron was also
included to show the behavior of an uncorrelated element. The
correlations involving Sr, Sc and Na are spurious, however, for the
following reasons :

\begin{figure}[htbp]
    \subfigure{\resizebox{\hsize}{!}{\includegraphics{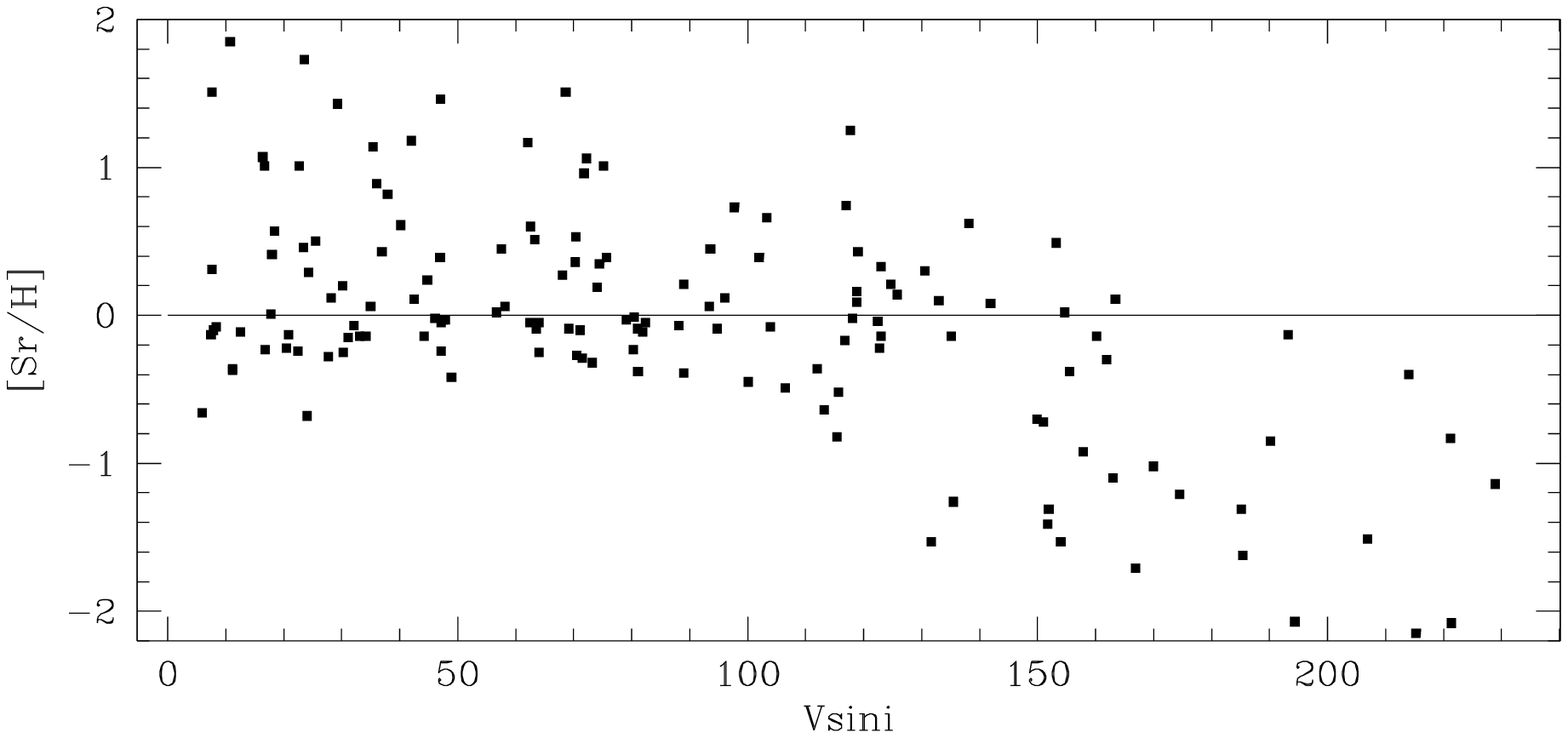}}}
    \subfigure{\resizebox{\hsize}{!}{\includegraphics{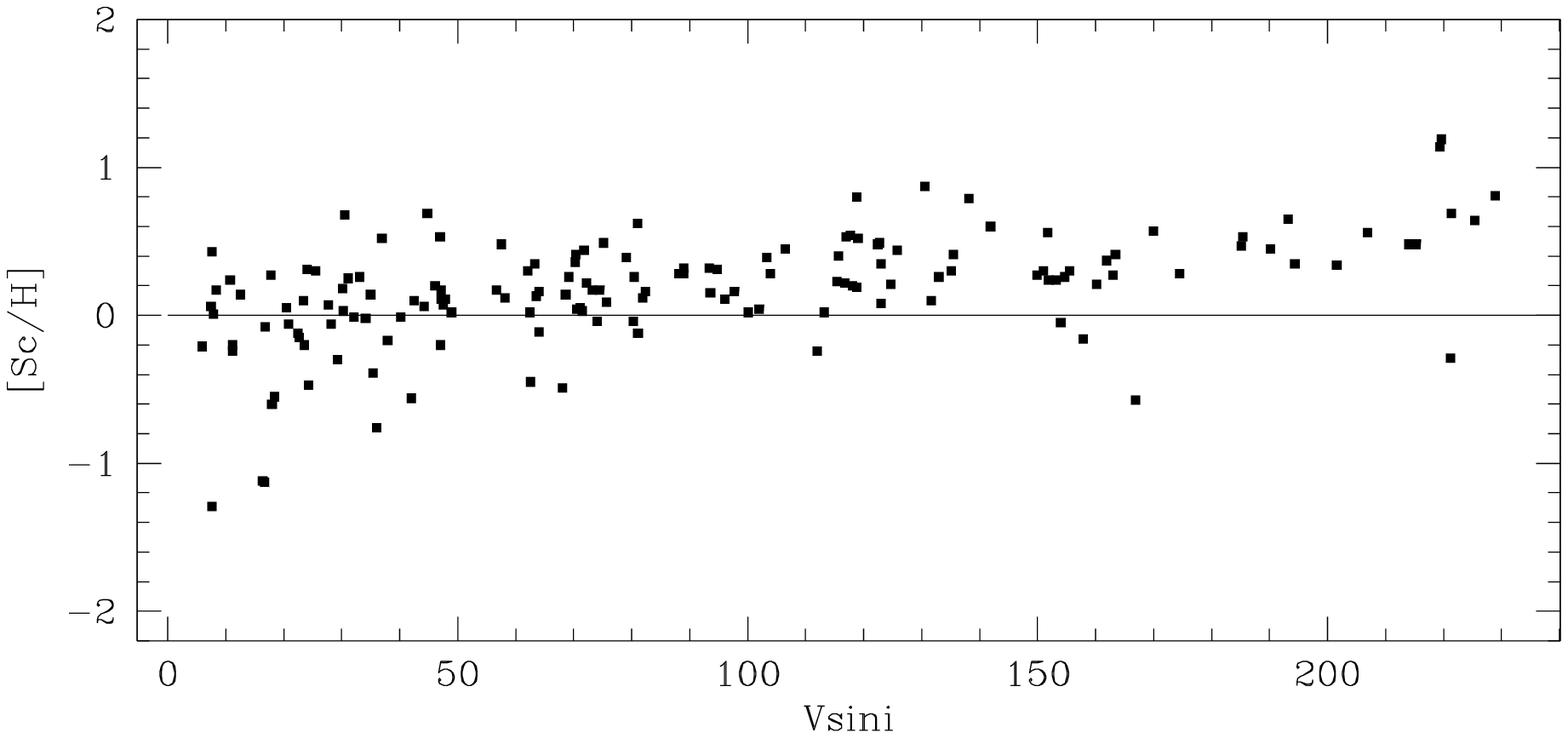}}}
    \subfigure{\resizebox{\hsize}{!}{\includegraphics{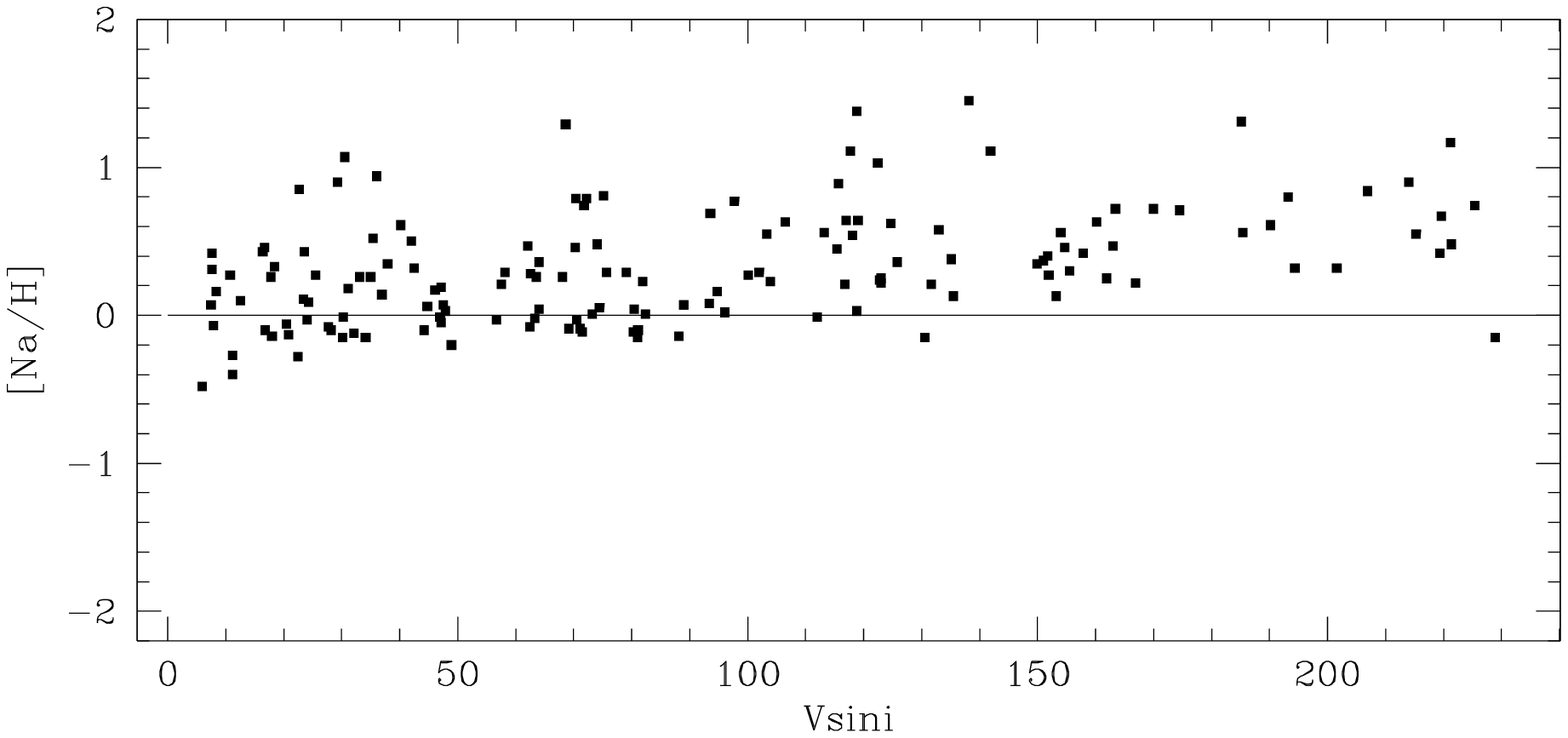}}}
    \subfigure{\resizebox{\hsize}{!}{\includegraphics{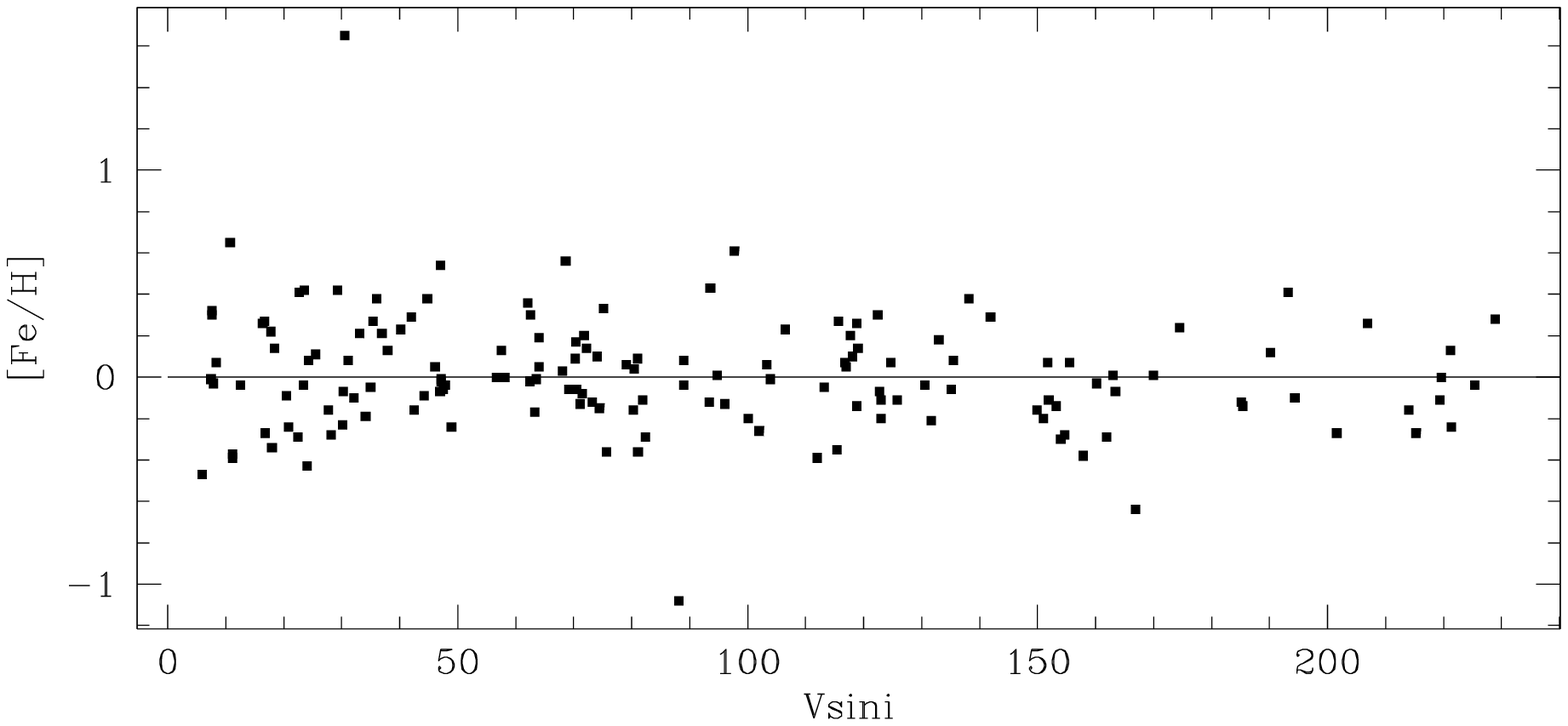}}}
\caption{Behaviour of Sr, Sc, Na, Fe with respect to \vsini.}
\label{vs}
\end{figure}

\paragraph{Strontium:}

This element shows the most significant correlation (correlation coefficient
$r= -0.61$). The problem arises when only 
one line, \ion{Sr}{ii} $\lambda$4215.519, is visible. This happens
mainly in the hotter stars of the sample. The fact is that 
the continuum is slightly underestimated in this region. When
rotational velocity increases, this small error leads to large
underestimates of the abundances. As there is no other line of Sr for
those stars, this error cannot be corrected. In Fig.~\ref{vs}, almost all stars
with \vsini$> 150$~\kms are of spectral type A, which reflects the well-known
drop of rotational velocity for spectral types F and later. It is for these
relatively hot stars that only one \ion{Sr}{ii} line is used, and in
this case, the result is much poorer than in the simulation
illustrated by Fig.~\ref{contfig}, 
which involves an F star where two Sr lines are used.

This is the only negative correlation for all adjusted elements. It is 
very interesting to illustrate the accuracy of the continuum
adjustment, and the 
stability of this method. Indeed, an underestimate of the continuum
leads to a negative correlation. The fact that only one element shows
such a behaviour is a good justification for our choice of continuum
adjustment.  
 
\paragraph{Scandium:}
This element yields the largest positive correlation ($r=0.5$). It should be
remembered that during the analysis, 
spectra are cut into 7 parts, each part is adjusted, and finally the
different estimates are put together (see Paper I for details). In
the case of Sc, there is a large 
scatter between abundances of different spectral ranges. Moreover,
since Sc is ionized, the adjustment is very sensitive to gravity (see
Sect.~\ref{param}). However, there is no correlation between \lgg and
\vsini. Unfortunately, the reason of this behaviour was not found, but
it is probable that there are several causes adding up.

\paragraph{Sodium:}
The last significant correlation ($r=0.43$) is found for Na. For this
element, the reason was found to be the telluric lines. The Na I doublet
($\lambda$ 5889 and $\lambda$ 5895)
lies within telluric lines. Even if the adjustment is not very
sensitive to these lines, a small error can lead to a large abundance
overestimate, especially for fast rotators, because these lines are
broad and saturated.

There is no significant correlation between other elements and
\vsini. 

\subsubsection{Correlation between rotation and true abundances}

\begin{figure*}[htbp]
\resizebox{\hsize}{!}{\includegraphics{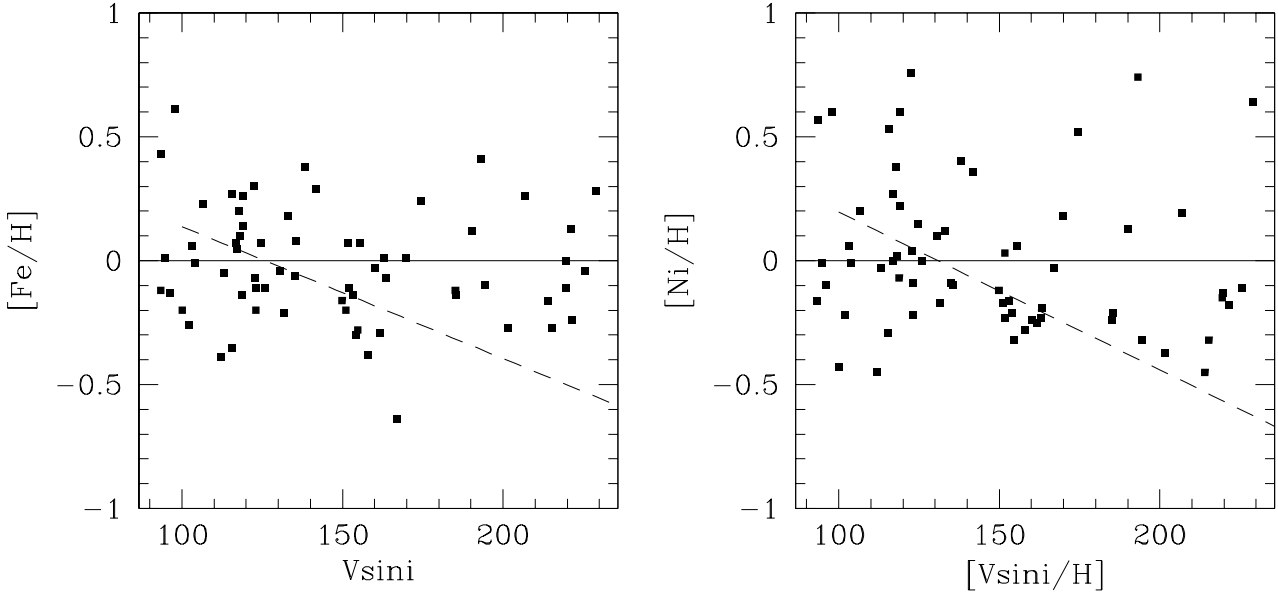}}
\caption{Abundances of Fe and Ni as a function of \vsini for fast
  rotators. The dashed line represent the anticorrelation proposed by
  Varenne \& Monier (\cite{varenne99}).}
\label{varenne}
\end{figure*}

The decrease of $\left[\frac{Fe}{H}\right]$ with increasing  \vsini
found  by Takeda \& Sadakane (\cite{takeda97}) and confirmed by
Varenne \& Monier (\cite{varenne99}) is firmly denied by our results. 
Fig.~\ref{varenne} shows clearly that there is no dependence of iron or
nickel abundance with \vsini. In our case, there are 60 stars with \vsini
$\geq$ 100 \kms while there was only 10 in Varenne \& Monier and 5 in
Takeda and Sadakane. It is important to notice that the
anticorrelation was found for stars of the Hyades, but, a priori, there is no
reason why only these stars should be affected. The result of Takeda et al. 
(\cite{takeda97}) is discussed in more details in Sect.~\ref{abund}

\subsection{Metallic giants}

\begin{figure}[htbp]
\resizebox{\hsize}{!}{\includegraphics{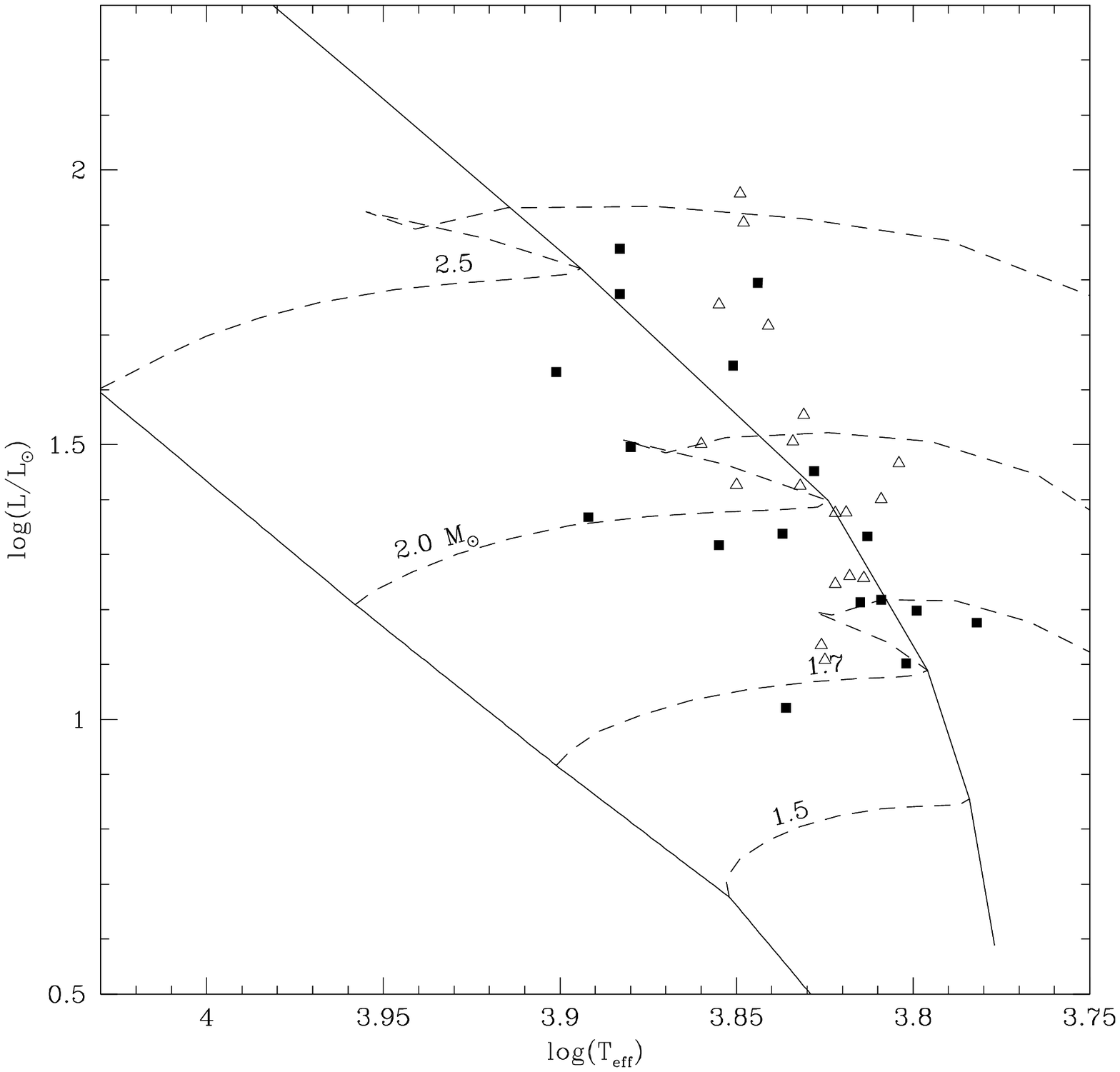}}
\caption{HR diagram of giant stars with some some evolutionary
  tracks (Schaller et al., \cite{schaller}). $\blacksquare$ represent
  normal giants with $\Delta\mathrm{m_2} < 
  0.013$ and $\triangle$ represent metallic giants with $\Delta\mathrm{m_2} \geq
  0.013$.}
\label{hr_iii}
\end{figure}

\begin{table}[htbp]
  \begin{center}
    \caption{Characteristics of the observed metallic giants.}
    \begin{tabular}{c|c|c|c|c}

HD   &  \teff  & \lgg  & \vsini  & spectral type  \\
\hline
432      &  6792  &   3.59    &   74.47  &F2 III\\
12311    &  7079  &   3.67    &  155.54  &F0 V\\
15233    &  6698  &   3.81    &  130.54  &F2 III\\
79940    &  6367  &   3.42    &  118.82  &F5 III\\
115604   &  7046  &   3.27    &    7.61  &F3 III\\
147547   &  7063  &   3.23    &  151.77  &A9 III\\
155203   &  6516  &   3.65    &  161.86  &F2 V (Cr)\\
157919   &  6576  &   3.67    &   46.96  &F3 III\\
159876   &  7244  &   3.65    &   23.41  &F0 IIIb ((Sr))\\
172748   &  6776  &   3.47    &   25.51  &F2 III*\\
181333   &  7161  &   3.41    &   47.83  &F0 III\\
186155   &  6637  &   3.70    &   44.76  &F5 II-III\\
196524   &  6441  &   3.50    &   46.10  &F6 III\\
200723   &  6591  &   3.57    &   93.42  &F2 III\\
205852   &  6934  &   3.38    &  163.43  &F3 III(a)\\
208741   &  6683  &   3.82    &   36.94  &F3-5 III\\
214441   &  6823  &   3.53    &  122.73  &F1 III\\
214470   &  6637  &   3.59    &   81.00  &F3 III-IV\\
\hline
       \end{tabular}
    \label{tab:iiip}
  \end{center}
\end{table}

\begin{table}[htbp]
  \begin{center}
    \caption{Characteristics of the observed normal giants.}
    \begin{tabular}{c|c|c|c|c}

HD   &  \teff  & \lgg  & \vsini  & spectral type \\
\hline
1671      &  6441  &   3.66    &   44.20 &F5 III\\
2628      &  7161  &   3.78    &   20.84 &A7 III\\
12573     &  7961  &   3.72    &  118.09 &A5 III\\
118216    &  6531  &   3.69    &   16.82 &F5 III\\
127762    &  7585  &   3.74    &  123.02 &A7 III\\
142357    &  6501  &   3.58    &   30.11 &F5 II-III\\
150557    &  6854  &   3.93    &   63.26 &F2 III\\
151769    &  6295  &   3.63    &   12.48 &F7 III\\
159561    &  7798  &   3.89    &  214.04 &A5 III\\
166230    &  7638  &   3.46    &   42.54 &A8 III\\
176303    &  6053  &   3.56    &   27.69 &F8 III\\
181623    &  6729  &   3.55    &  112.03 &F2 III\\
182900    &  6338  &   3.73    &   31.08 &F6 III\\
186005    &  6870  &   3.69    &  151.02 &F1 III\\
187764    &  6982  &   3.33    &  102.00 &F0 III\\
197051    &  7638  &   3.53    &   75.68 &A7 III\\
208450    &  7095  &   3.49    &  131.66 &F0 III-IVn\\
\hline
       \end{tabular}

    \label{tab:iiim}
  \end{center}
\end{table}

Fig.~\ref{hr_iii} shows the position of all giant stars of our sample in the HR
diagram. There are 35 giants, out of which 18 are metallic. Their
characteristics are reported in Tables~\ref{tab:iiip} and
\ref{tab:iiim}.

\begin{figure*}[htbp]
\resizebox{\hsize}{!}{\includegraphics{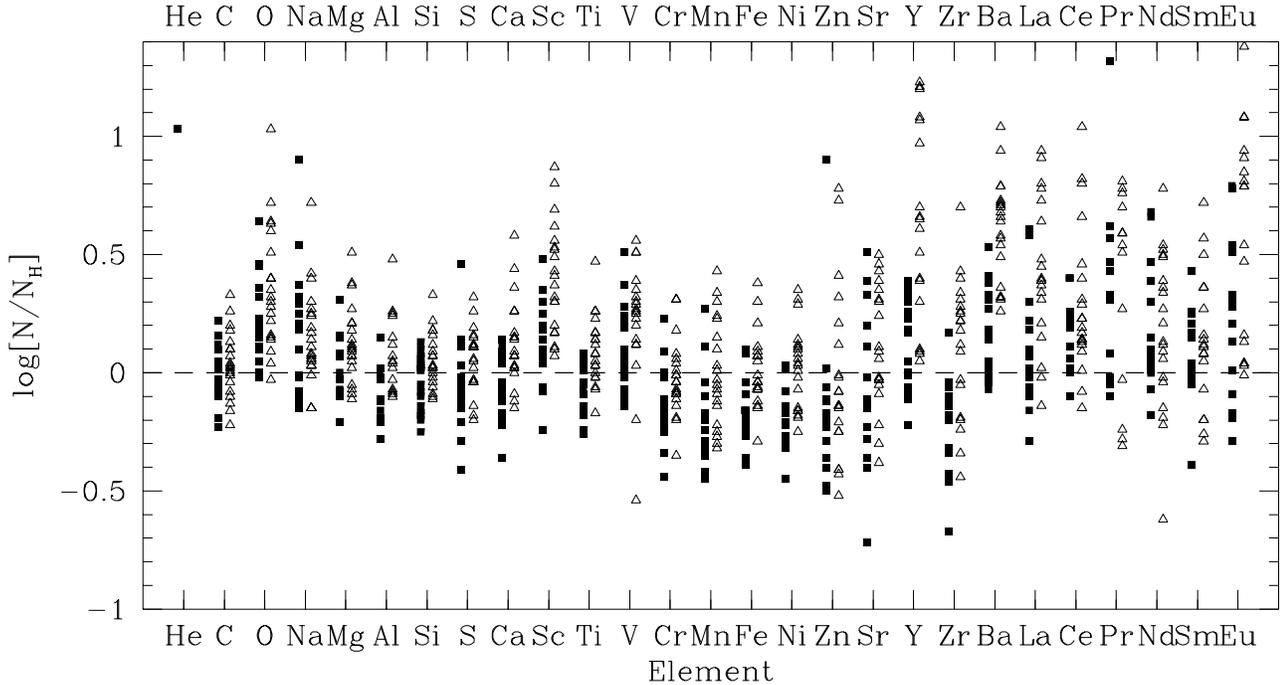}}
\caption{Abundance pattern of normal giants ($\blacksquare$) and of
  metallic giants ($\triangle$)}
\label{compiiipm}
\end{figure*}

Abundances of observed giants are illustrated in
Fig.~\ref{compiiipm}. There are no very clear-cut peculiarities such as the
underabundance of Ca and/or Sc in Am stars. However, abundances of
some elements (Al, Ca, Ti, Fe, Ba for example) are clearly different.

To ensure that the abundances are really different, the Kolmogorov-Smirnov 
test was used. This test gives the probability that two distributions
are drawn from the same parent population. 

It was already shown in North et al. (\cite{north98}) that the \lgg
distribution of metallic and normal giants are different with a
probability of 99.99\%. However, the \teff distributions are comparable. 
The distributions of \vsini and \vmic estimated by our method are also
comparable.

\begin{figure*}[htbp]
  \mbox{
 \hspace{-.5cm}
        \subfigure{\includegraphics[width=6.cm]{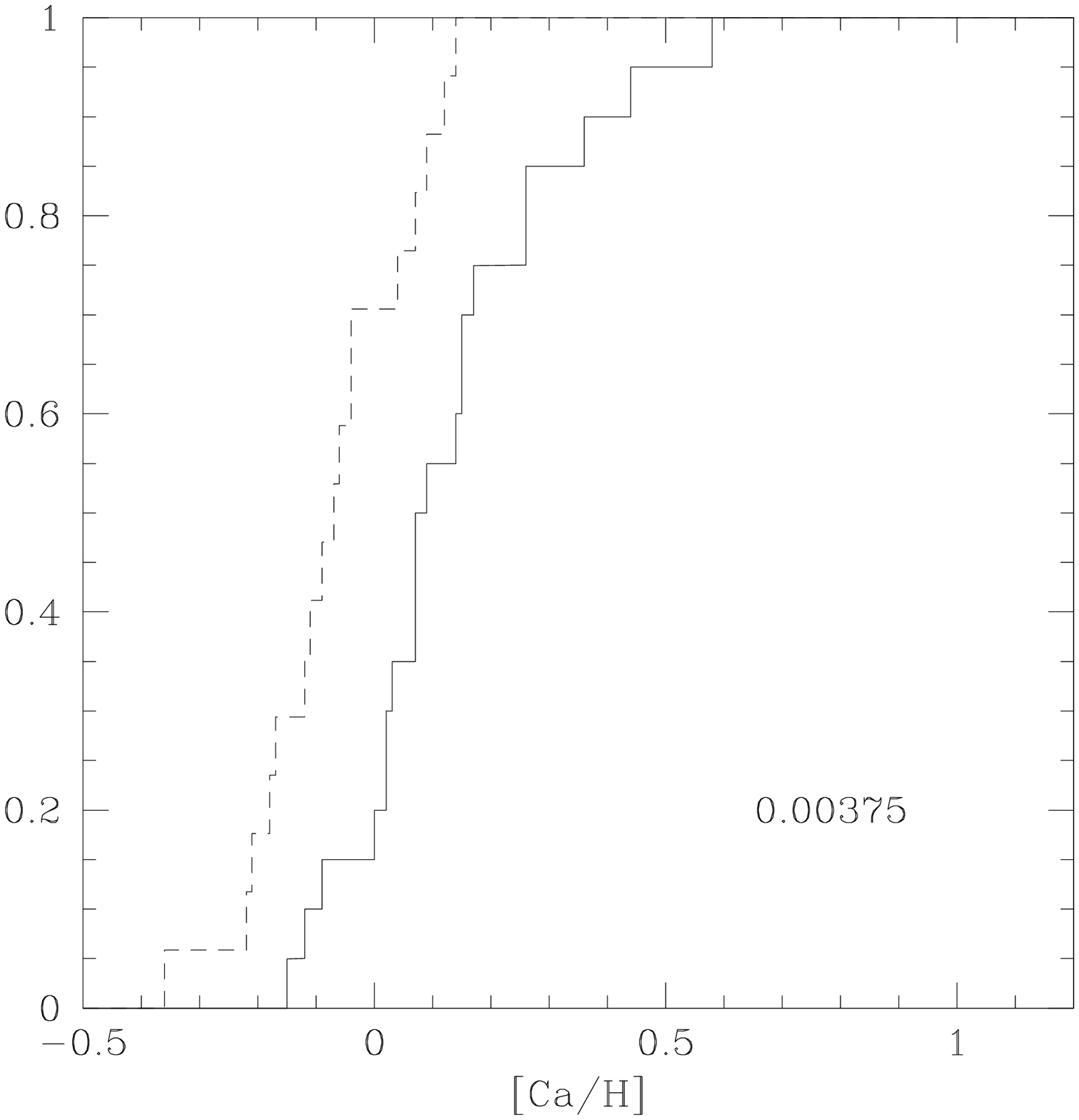}}
        \subfigure{\includegraphics[width=6.cm]{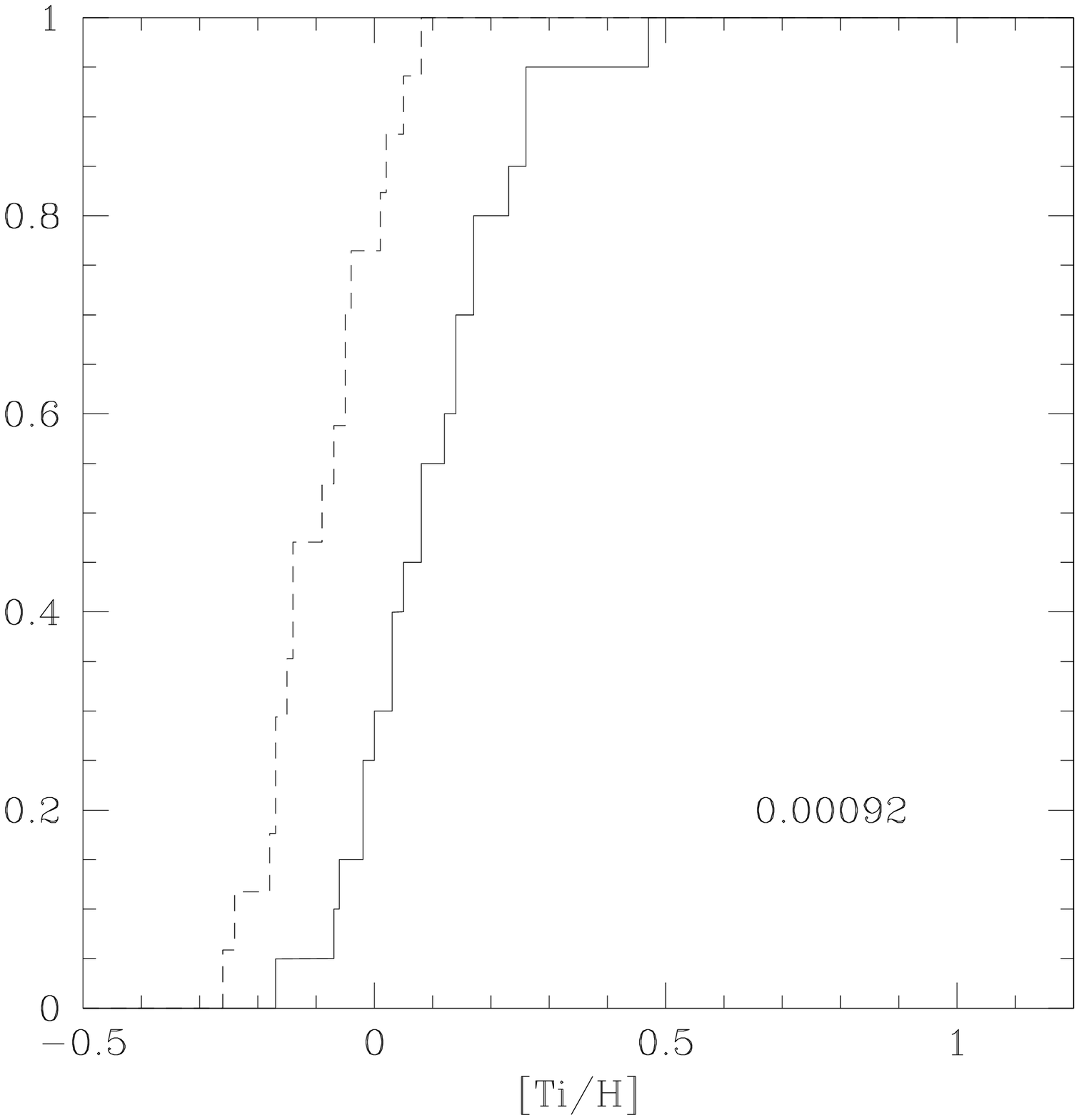}}
        \subfigure{\includegraphics[width=6.cm]{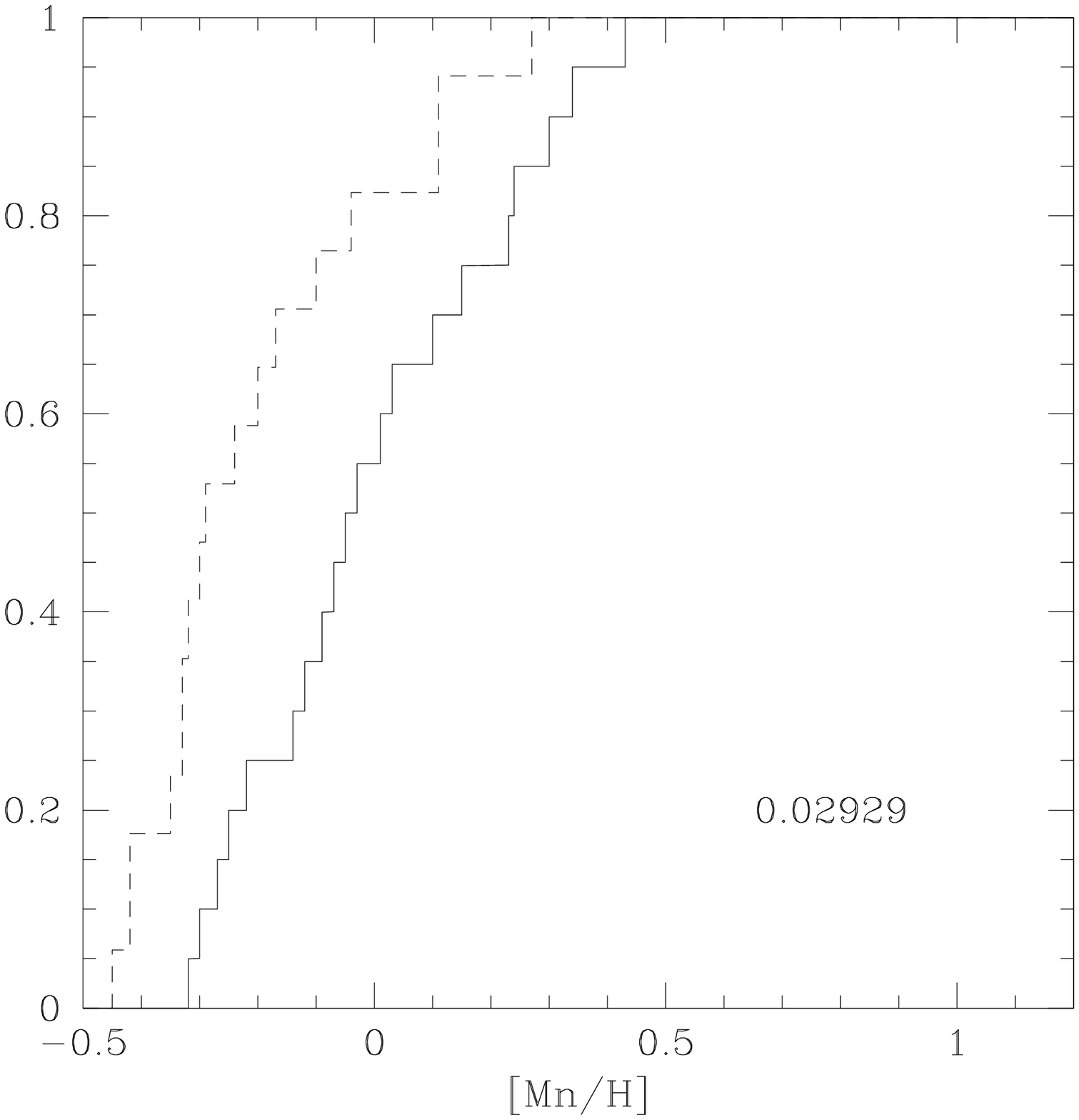}}
  }
 \mbox{
 \hspace{-.5cm}
        \subfigure{\includegraphics[width=6.cm]{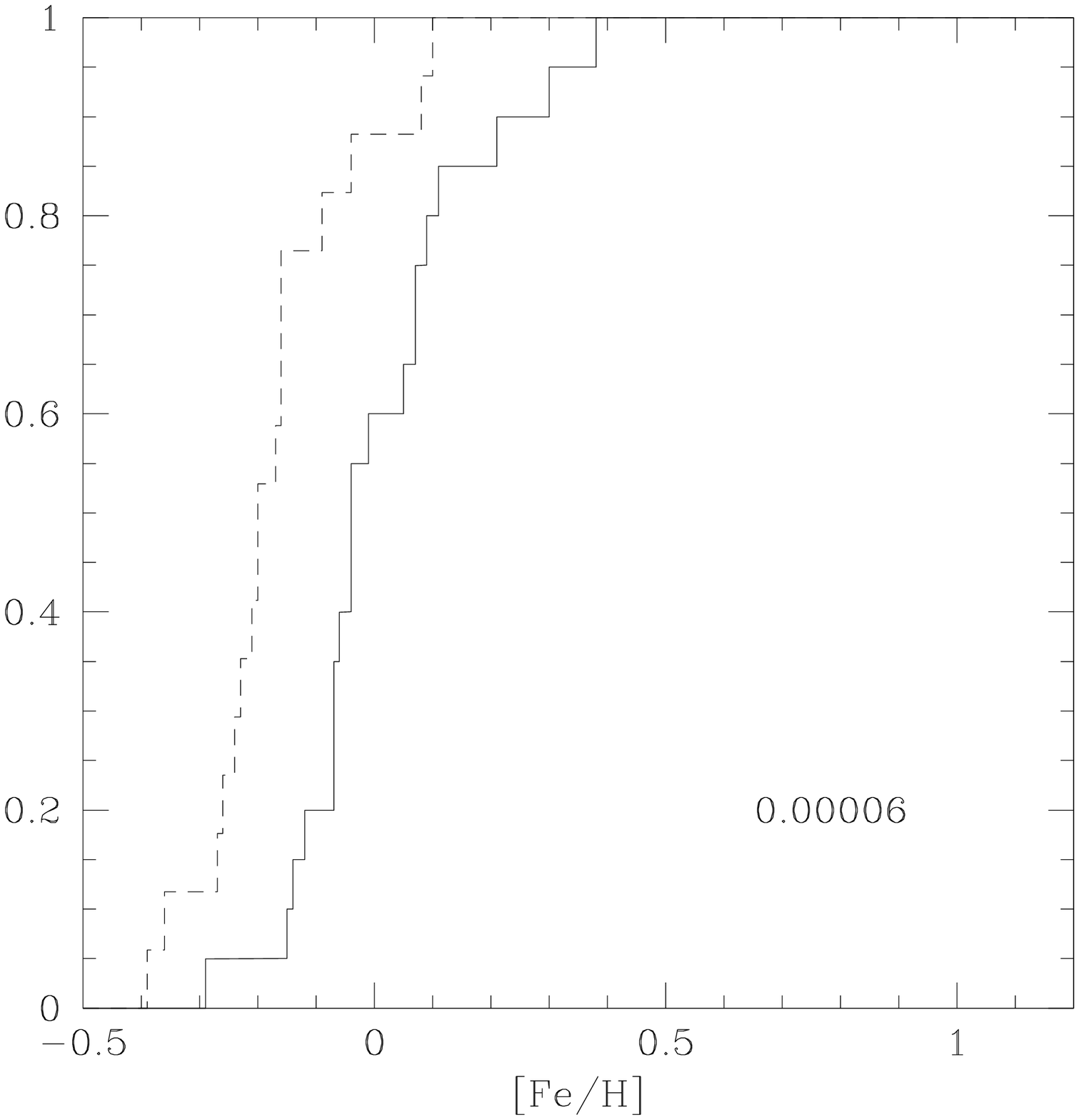}}
        \subfigure{\includegraphics[width=6.cm]{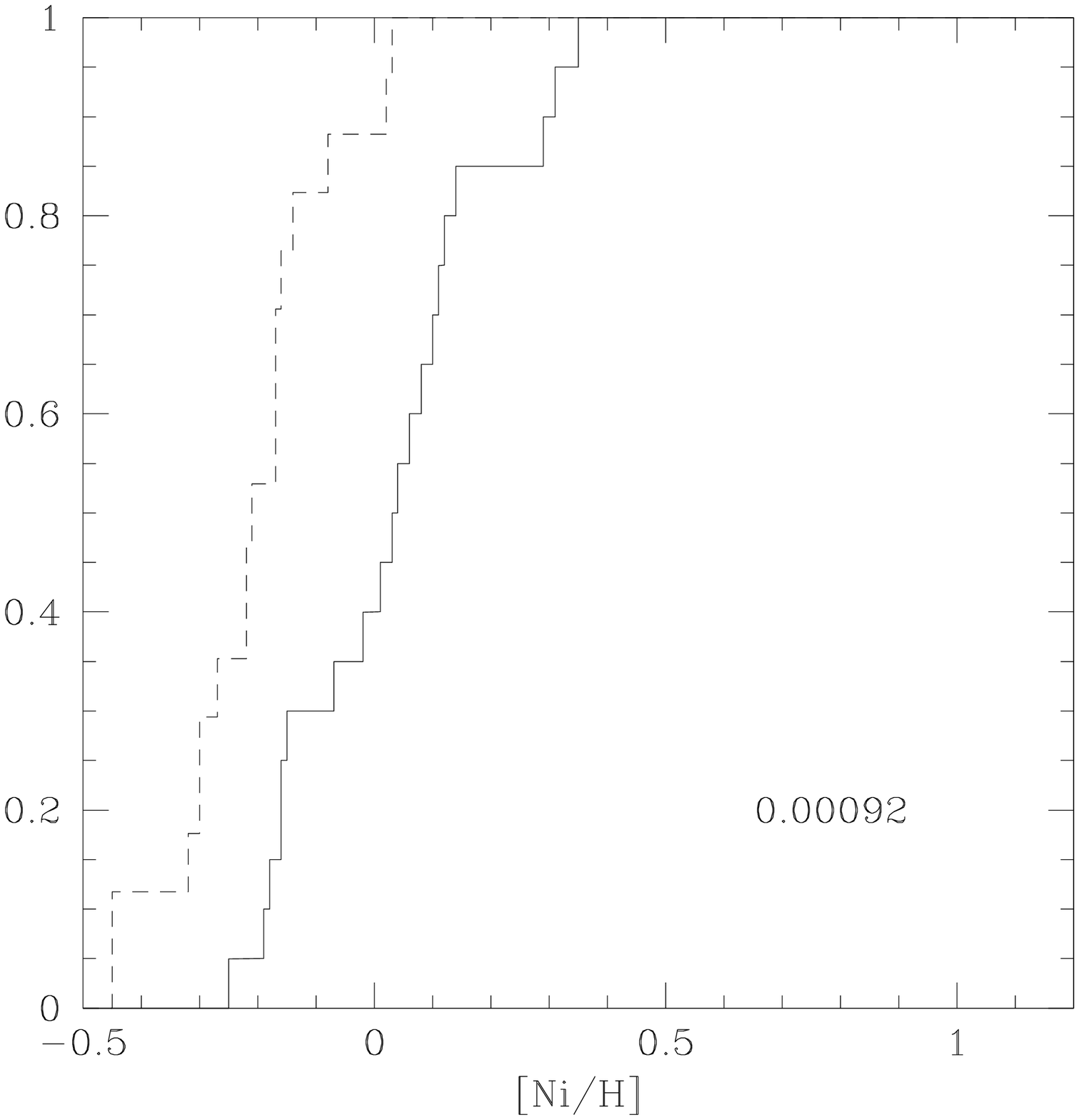}}
        \subfigure{\includegraphics[width=6.cm]{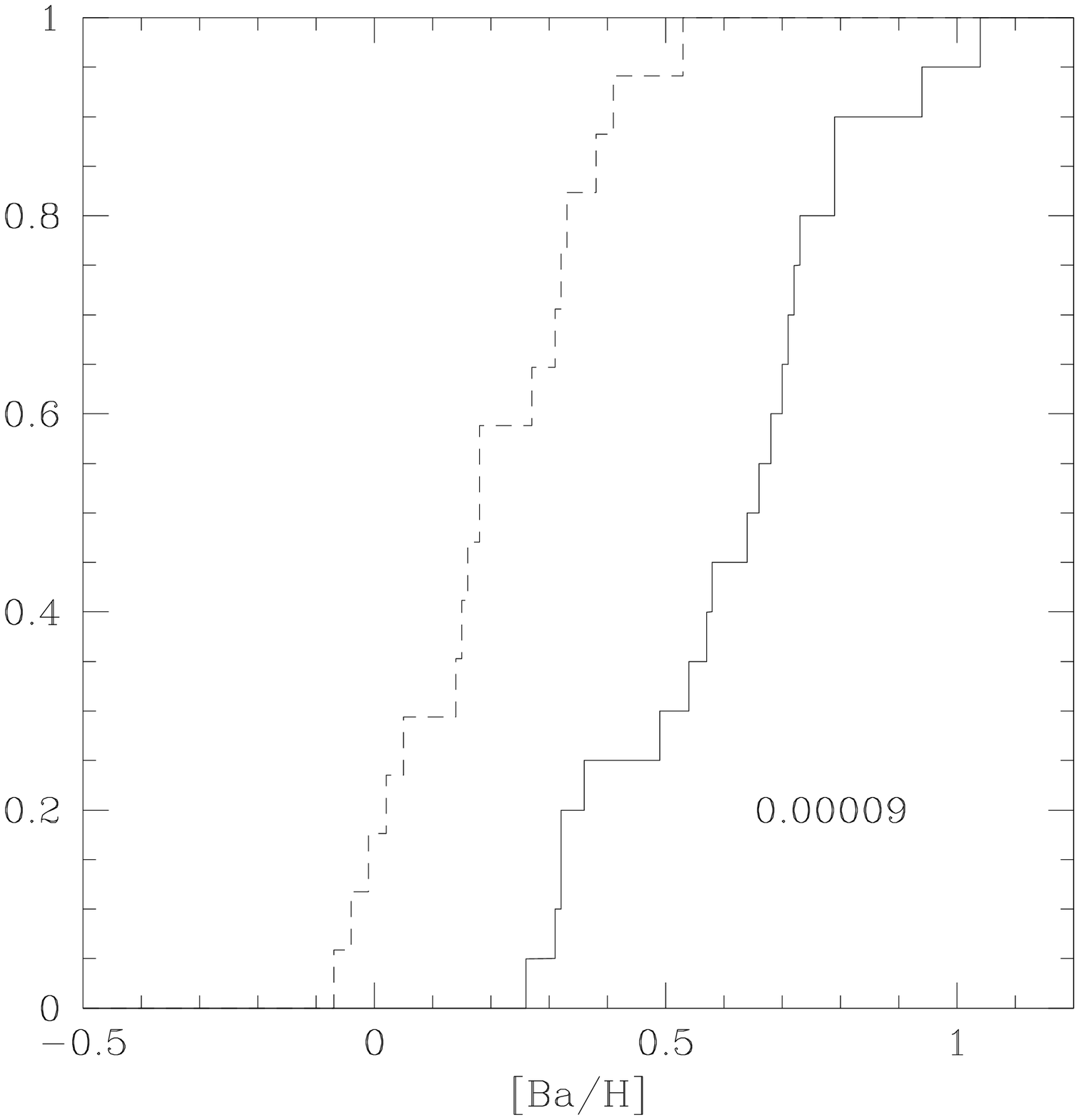}}
  }
  \caption{Cumulative distributions of abundances of metallic giants
    (solid line) and of normal one (dashed line). The number in each
    panel indicates the probability that 
    both distributions are drawn from the same parent distribution.}
  \label{smiriii}
\end{figure*}

Fig.~\ref{smiriii} shows the cumulative distribution of abundances for
each of the 6 elements that have 
the largest differences. They are Ca, Ti, Mn, Fe, Ni and Ba. Al and Cr
also have different distributions. Sc is not added, even though the
corresponding 
distributions are clearly different, because of its correlation with
\vsini.

Thus, we conclude that metallic giants have abundances of Al, 
Ca, Ti, Cr, Mn, Fe, Ni and Ba that are significantly larger than those
of normal giants. 

\subsection{Abundance variation with evolution}
\label{abund}

\begin{figure*}[htbp]
\resizebox{\hsize}{!}{\includegraphics{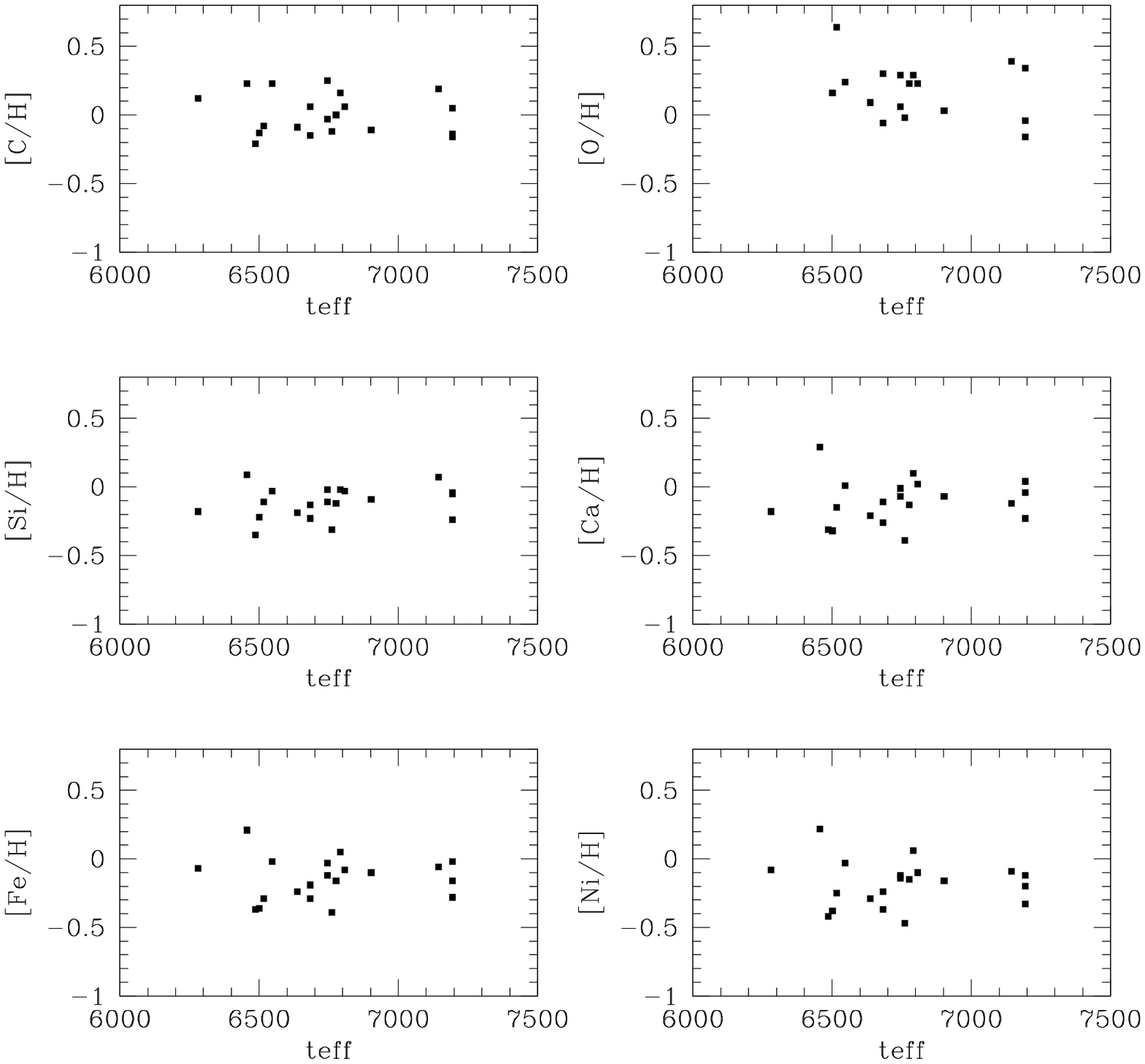}}
\caption{Abundances of dwarf F-type stars as a function of \teff. }
\label{varenne1}
\end{figure*}

\begin{figure*}[htbp]
\resizebox{\hsize}{!}{\includegraphics{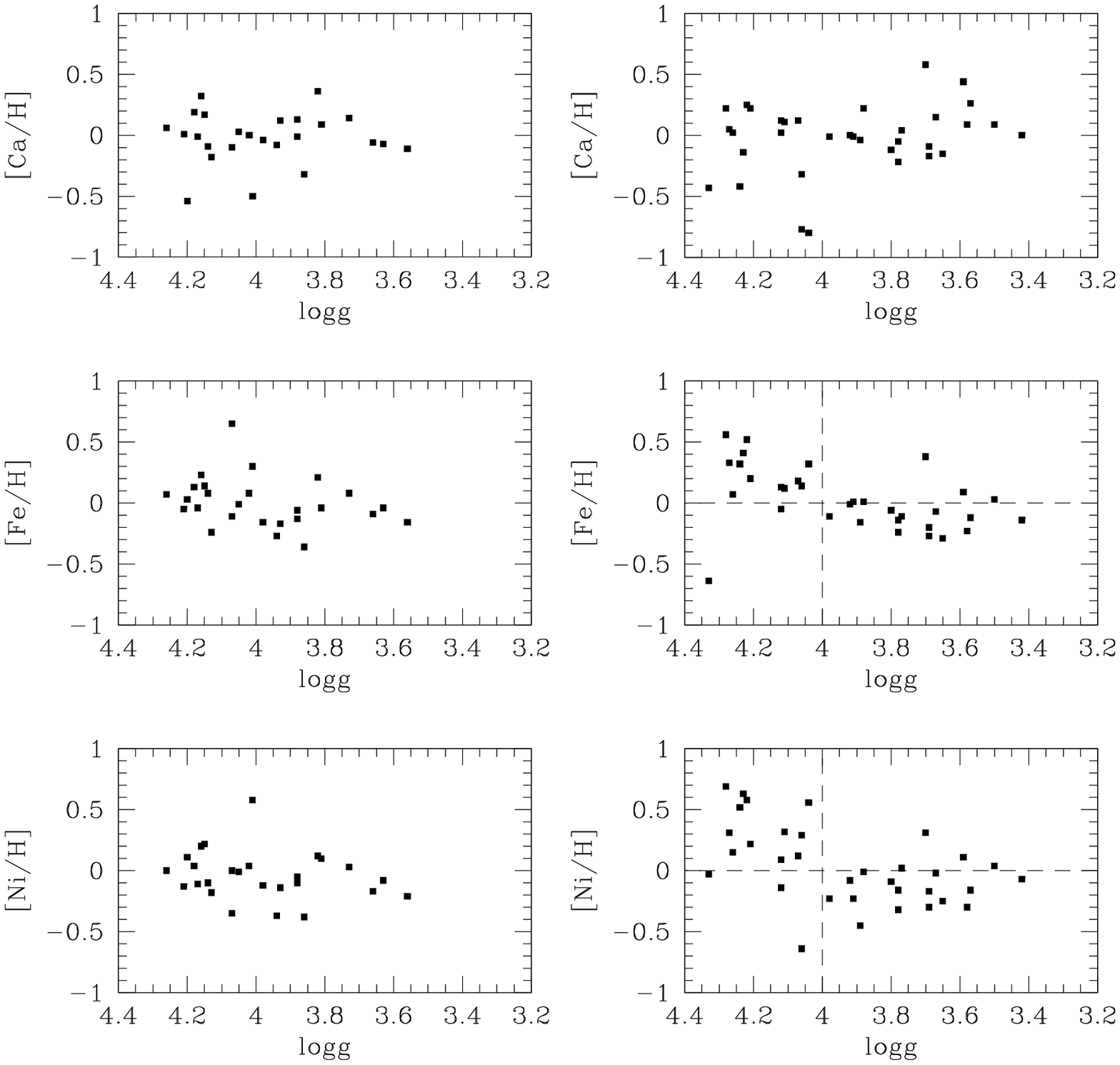}}
  \caption{Relation between abundances of Ca, Fe and Ni with
    \lgg. \textbf{Left:} stars of 1.6 to 1.8 \msol. \textbf{Right:}
    stars of 1.8 to 2.0 \msol. The broken lines show the limits of the
    $2\times 2$ contingency tables mentioned in the text.}
  \label{richard}
\end{figure*}

Recently, the Montr\'eal group has published a new study for normal
stars. Models including all effects of atomic diffusion and radiative
accelerations have been presented for stars of different masses in Turcotte et
al. (\cite{turcotte98}), Richer et al. (\cite{richer00}) and Richard
et al. (\cite{richard01}). 

The models of Turcotte et al. predict the evolution of
the abundances for an F star of a   
given mass consistently with the internal structure. They are computed
in the diffusion framework with 
only the mixing length as free-parameter. They include the
effect of 28 chemical elements (Z $\leq$ 28) but no macroscopic
mixing. Atomic diffusion plays an important role for stars more
massive than 1.3 \msol.
We agree with Varenne \& Monier (\cite{varenne99}) that the theoretical prediction of 
Turcotte et al. (\cite{turcotte98}) for F stars are not confirmed by
observations.   In our case, abundance of C,
O, Si, Ca, Fe, and Ni do not show any variations with effective
temperature (see Fig.~\ref{varenne1}).

Richard et al. (\cite{richard01})
present stellar models
that are evolved from the pre-main sequence to the giant branch
for stars of 1.3 to 4.0 \msol and metallicities of Z=0.01 to 0.03.
In order to compare our observations with these models, our sample was
split into 7 mass groups with
 masses below 1.4 \msol, 1.4 to
1.6 \msol, 1.6 to 1.8 \msol, \ldots, 2.4 to 2.6 \msol and above
2.6 \msol.
Fig.~\ref{richard} shows the behaviour of Ca, Fe and Ni for stars of
1.6 to 1.8 \msol and 1.8 to 2.0 \msol with respect to \lgg. 
It can be directly compared with Fig.~10 of Richard et
al. (\cite{richard01}). They predict a large abundance variation of at
least 1 dex with decreasing \lgg (or equivalently, with time ). In our
case, there is no variation with \lgg except for stars of 1.8 to 2.0
\msol, and even in this case, the 
amplitude is much smaller than the theoretical predictions.
In order to confirm the statistical significance of the correlation, $2\times 2$
contingency tables have been computed, segregating stars with gravities larger
or lower than \lgg$=4.0$ and with [Fe/H] or [Ni/H] larger or lower than $0.0$.
The corresponding $\chi^2$ were computed taking Yates' correction into account
(Crow et al. \cite{CDM60}) and the results were $8.98$ and $10.94$ for [Fe/H]
and [Ni/H] respectively, which comfortably exceeds the $7.88$ value
corresponding to a 0.5\% probability that abundances and \lgg are unrelated.
Thus, taken at face value, the correlation is statistically significant.
If this variation is physically true, it suggests that additional hydrodynamical
processes are at play in 
these stars, which lower the effects of microscopic diffusion.
However, before concluding to the physical reality of this correlation, 
it is necessary to test how far this behaviour can be explained by a
correlation of abundances with \vsini or by NLTE effect.

Takeda \& Sadakane
(\cite{takeda97}) also found a variation of iron abundance with
\lgg, which was qualitatively the same, for Hyades' members. They
explain this variation as a superficial phenomenon due to 
the existence of a mutual relationship between \lgg and \vsini. They
found an anticorrelation between \vsini and \lgg, in the sense that
high-gravity stars tend to be slow rotators. They consider that the
cause of this relation is rotation, by which the apparent and effective 
values of \lgg are affected. In their case, \lgg is estimated from Str\"omgren
photometry, and rotation affects this determination:
it appears that \lgg values progressively decrease
with increasing \vsini. 

\begin{figure}[htbp]
\resizebox{\hsize}{!}{\includegraphics{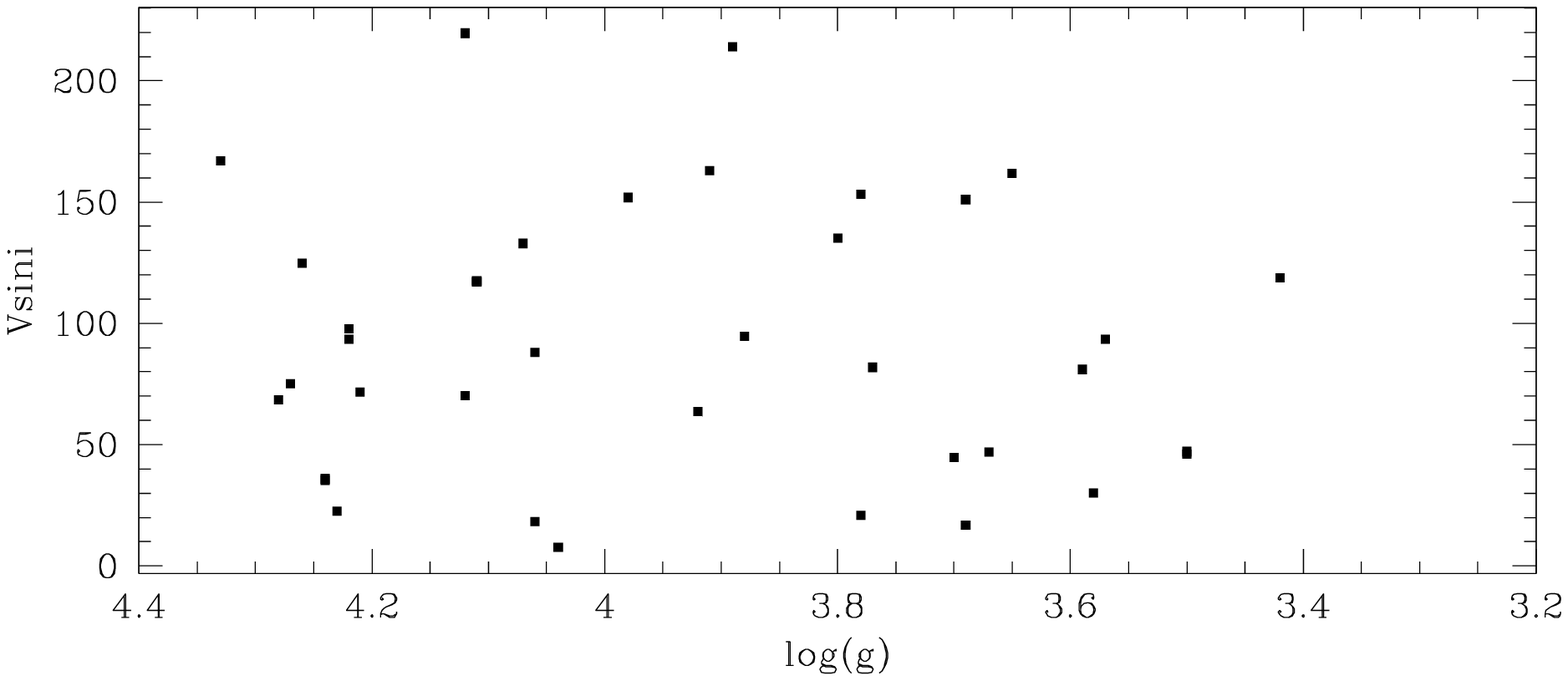}}
  \caption{Plot showing \vsini as a function of \lgg for stars of 1.8
    to 2.0 \msol.}
  \label{vs_logg}
\end{figure}

However, in our case, \lgg values were estimated from \textsc{Hipparcos}
data, and they are not correlated with \vsini (see
Fig.~\ref{vs_logg}).
The effect of rotation is to increase the $B2-V1$ index.
  Maeder (\cite{maeder}) has shown that, to a good approximation, this
  index is essentially sensitive to the product \vsini. For a star of
  1.4 \msol (2.0 \msol
  respectively), a \vsini of 200 \kms leads to a decrease in  
  log(\teff) of 0.016 (0.010 respectively). The effect on $M_V$
  is of a few hundredths of a magnitude ($\sim 0.1$~mag respectively)
  (Maeder \& Peytremann, \cite{maeder70}). This affects slightly the mass
  estimate when interpolating through evolutionary tracks of non-rotating
  models, but the error cannot exceed 0.1 \msol (5\% at 2 \msol), propagating
  into no more than $\sim 0.02$~dex in \lgg. In fact our estimates of \lgg
  are close to the effective gravity, i.e. the sum of gravitational and
  centrifugal accelerations average over the visible disk; we verified that
  they agree quite well with photometric estimates computed from the
  $uvby\beta$ colours and the calibration by Moon \& Dworetsky (\cite{MD85}).
  The interesting thing with effective \lgg is that for \vsini$\sim
  200$~\kms, it differs by no more, on the equator, than 0.1~dex from a
  non-rotating 2~\msol star near the ZAMS, while the difference is at least
  twice as large at the turn-off. Thus the observed \lgg axis (e.g. in
  Fig.~\ref{richard}) is slightly stretched for fast rotators, relative to
  the \lgg non-rotating stars would have in the same mass range. But since slow
  rotators are not affected, the net effect is just to blur very slightly the
  relation metallicity vs. \lgg shown in Fig.~\ref{richard}.
Therefore, this cannot explain the behaviour of 
$\left[\frac{Fe}{H}\right]$ with \lgg. Moreover, in 
Takeda \& Sadakane (\cite{takeda97}), \lgg varies on a range that is
no more than 0.2 dex; under the assumption that this variation is due entirely
to a bias in the \lgg estimate, it can hardly explain a 0.4 dex variation of
$\left[\frac{Fe}{H}\right]$ for the single line of neutral iron they used. 

Given the variation of \teff and \lgg for these stars, NLTE effects
might play an important role. Rentzsch-Holm (\cite{rentzsch}) has
estimated NLTE effects on iron abundances for A-type stars. NLTE
correction increase with \teff going from 7\,000 to 10\,000 K (see
Fig.~4 of  Rentzsch-Holm 
(\cite{rentzsch})). On the other hand, at given effective temperature,
they increase with decreasing gravity. Since both \teff and \lgg
decrease with time when a star is evolving on the main sequence, this tends to
flatten the NLTE correction. For our stars of 1.8 to 2.0 \msol, there is a 
small NLTE correction of 0.1 to 0.15 dex. However, this correction is
almost constant (variation of 0.05 dex) for the \teff and \lgg
considered\footnote{Note that this is valid only if \teff and \lgg are
  correlated. The Pearson product-moment correlation coefficient is
  0.972 (0.968) for stars of 1.8 to 2.0 \msol (1.6 to 1.8 \msol
  respectively), and in the same order for the other mass interval.}.
Moreover, with our method, we do not have
separate abundance values for \ion{Fe}{i} and \ion{Fe}{ii}. 
Let us recall that \ion{Fe}{ii} lines are almost not affected by NLTE effects.
When \teff increases, the ratio of
line numbers of \ion{Fe}{i} and \ion{Fe}{ii} decreases. Therefore our
determination becomes less sensitive to NLTE effects. This also tends
to flatten the variation of NLTE correction with \teff. Therefore, we
conclude that, for stars of 1.8 to 2.0 \msol, only a small NLTE
correction of about 0.1 dex, that is roughly
constant with \teff, has to be taken into account when iron abundance
is estimated with our method. To summarize, all points in the
$\left[\frac{Fe}{H}\right]$ vs \lgg diagram for stars of 1.8 - 2.0
\msol in Fig.~\ref{richard} should be shifted upward by about 0.1 to
obtain a more realistic picture.

Th\'evenin \& Idiart (\cite{thevenin99}) discuss the problem of \lgg
determination when dealing with NLTE correction. Their work focuses on 
cooler stars, but the problem of \lgg determination is
comparable. When the NLTE correction is applied to LTE abundances,
abundances of 
\ion{Fe}{i} and \ion{Fe}{ii} are no longer the same (assuming that
\lgg had been adjusted to satisfy the LTE ionization
equilibrium). Therefore, \lgg has  
to be corrected until ionization equilibrium is satisfied \textbf{after} NLTE
correction. Nissen et al. (\cite{nissen97}) have shown that there
exists a discrepancy between spectroscopic \lgg (adjusted assuming
LTE) and those deduced from 
\textsc{Hipparcos} parallaxes. 
Th\'evenin \& Idiart (\cite{thevenin99}) have solved this discrepancy
by adjusting \lgg after NLTE correction; the latter
 are in close agreement with those deduced from
\textsc{Hipparcos} (our \lgg are estimated in this way). However, in
order to apply NLTE 
corrections precisely, it is necessary to have separate abundance estimates of
\ion{Fe}{i} and \ion{Fe}{ii}, which is not our case. Roughly, our
determination can be seen as an average of both abundances weighted by 
the number of lines of each ionization stage. 

To summarize, NLTE effects on iron is no more than 0.1 dex and remains 
constant to better than 0.1 dex, for stars
of 1.8 to 2.0 \msol when abundance are estimated with our method.

In view of the above discussion, it seems that the correlation between 
$\left[\frac{Fe}{H}\right]$ and \lgg for 1.8-2.0 \msol is indeed
physical, and due neither to systematic errors nor to selection effects.
It agrees qualitatively with the prediction of Richard et
al. (\cite{richard01}), but it is strange that only the stars of 1.8
to 2.0 \msol follow their models. It is true that more massive stars
are less numerous, so their statistics is less good.
For less massive stars, models predict a variation that occurs on a shorter
\lgg range. It is possible that this variation is too quick to be
observed, but then we would expect to see at least an increase of the scatter
for large \lgg values, while this is not the case.

The comparison with the Montr\'eal models shows that it is necessary to add
hydrodynamical processes in order to reproduce observations, like
mass loss and meridional circulation, as already suggested by the Montr\'eal
group. Maybe turbulence will also have to be modelled in a slightly different
way yet to be defined.

We also note that the metal-rich 1.8 to 2.0 \msol stars lie just around the blue border
of the instability strip or outside of it: the coolest of these (HD 144197) has
\teff$=7691$~K and the next cooler one (HD 148367) has \teff$=7962$~K,
while the blue limit of the instability strip is defined by 
$B2-V1= -0.01$ on the ZAMS (Jasniewicz \cite{jas84}) which correspond 
to \teff$=8016$~K according to the calibration of Hauck \& K\"unzli
(\cite{HK96}).  
The hottest metal-rich
star (HD 195943) has \teff$=9057$~K. Almost all
stars with $[Fe/H] < 0.0$ are cooler and lie, therefore, either inside the
instability strip or to the red side of it. This might be more than a
coincidence, and is reminiscent of the well known exclusion between
$\delta$~Scuti type pulsations and the Am phenomenon. The difference is that
while most Am stars do lie inside the instability strip even though they do not
pulsate, the metal-rich 1.8 to 2.0 \msol stars are found either outside or on
the verge of it.

\section{Conclusion}
 
The method of analysis presented in Paper I was used to derive
detailed abundances of 140 A and F-type stars with \vsini up to about
200 \kms. 
\vsini derived with this method are consistent with those in the 
literature. Only Cr, Sc, and Na are very sensitive to rotational
velocity in the sense that their abundance determination is altered
when \vsini $\geq$ 100 \kms, for various reasons.

This study allows us to specify more exactly the nature of metallic A-F
giants. They are more evolved than normal giants and they show
overabundances of at least 8 elements with respect 
to normal giants. These elements are Al, Ca, Ti, Cr, Mn, Fe, Ni and
Ba. 

Our results do not confirm, in general, the predictions of the
Montr\'eal models. However, stars of 1.8 to 2.0 \msol show a clear
correlation of iron 
abundances with \lgg which is reminiscent of theoretical predictions
by Richard et al. (\cite{richard01}). Neither rotational velocity nor
NLTE effects can explain this behaviour, which appears, therefore, to
be real. The observed $\left[\frac{Fe}{H}\right]$ and
$\left[\frac{Ni}{H}\right]$ variations qualitatively agree with those
predicted by the Montr\'eal models, but are much weaker. The smaller amplitude
indicates that their models need more hydrodynamical processes (meridional
circulation, mass loss, \ldots) and/or a better tuning of convection.
Our large homogeneous set of abundances hopefully can help this work. 

\begin{acknowledgements}
We express our warm thanks to P. Bartholdi and C. Briner 
for their help in using the computer network of the Geneva observatory 
and for the instructive discussions.
Constructive criticism by the referee, Dr. Richard Gray, is gratefully
acknowledged.
This work has been partly supported by the Swiss National Science Foundation.
\end{acknowledgements}


\begin{thebibliography}{}
\bibitem[1995]{abt} Abt, H.A. \& Morell, N.I., 1995, ApJS, 99, 135
\bibitem[1990]{balach} Balachandran, S., 1990, ApJ, 354, 310
\bibitem[1996]{baranne} Baranne, A., Queloz, D., Mayor, M., Adrianzyk, G.,
  Knispel, G., Kohler, D., Lacroix, D., Meunier, J.-P., Rimbaud, G. \& Vin,
  A., 1996, A\&AS, 119, 373
\bibitem[1992]{berthet92} Berthet, S., 1992, A\&A, 253, 451
\bibitem[1991]{berthet91} Berthet, S., 1991, A\&A, 251, 171
\bibitem[1990]{berthet90} Berthet, S., 1990, A\&A, 227, 156
\bibitem[1979]{B79} Breger, M., 1979, PASP, 91, 5
\bibitem[1996]{castelli} Castelli, F., Gratton, R.G. \& Kurucz, R.L., 1996, 
  A\&A, 318, 841
\bibitem[2001]{cayrel} Cayrel de Strobel, G., Soubiran, C. \& Ralite, N.,
  2001, A\&A, 373, 159
\bibitem[1993]{charbonnel} Charbonnel, C., Meynet, G., Maeder, A., Schaller, G.
\& Schaerer, D., 1993, A\&AS, 101, 415 
\bibitem[1992]{coupry92} Coupry, M.F. \& Burkhart, C., 1992, A\&AS, 95, 41
\bibitem[1999]{C99} Cramer, N., New Astron. Rev., 43, 343
\bibitem[1975]{C75} Crawford, D.L., AJ, 80, 955
\bibitem[1979]{C79} Crawford, D.L., AJ, 84, 1858
\bibitem[1960]{CDM60} Crow, E.L., Davis, F.A. \& Maxfield, M.W., 1960,
  Statistics Manual, Dover, New York
\bibitem[2002]{erspamer} Erspamer, D., North, P., 2002, A\&A, 383, 227
  (Paper I)
\bibitem[1997]{esa} ESA, 1997, The \textsc{Hipparcos} and Tycho
  Catalogues, ESA SP-1200 
\bibitem[1996]{flower} Flower, P.J., 1996, ApJ, 469, 355
\bibitem[1980]{G80} Golay, M., 1980, Vistas Astron., 24, 141
\bibitem[2001]{gray} Gray, R.O., Graham, P.W. \& Hoyt, S.R., 2001, ApJ, 121, 
  2159
\bibitem[1989a]{gray89a} Gray, R.O., Garrison, R.F., 1989a, ApJS, 70, 623
\bibitem[1989b]{gray89} Gray, R.O., Garrison, R.F., 1989b, ApJS, 69, 301
\bibitem[1987]{gray87} Gray, R.O., Garrison, R.F., 1987, ApJS, 65, 581G
\bibitem[1986]{hauck} Hauck, B., 1986, A\&A, 155,371
\bibitem[1973]{hauck73} Hauck, B., 1973, in IAU Symp 54, Vol 54, pp. 117
\bibitem[1996]{HK96} Hauck, B. \& K\"unzli, M., 1996, Baltic Astronomy, 5, 303
\bibitem[1991]{bsc} Hoffleit, D, Jaschek, C., 1991, \textit{The
    bright star catalogue}, 5th revised edition (Yale Univ. Obs., New Haven)
\bibitem[1984]{jas84} Jasniewicz, G., 1984, A\&A, 141, 116
\bibitem[1998]{kunzli98} K\"unzli, M. \& North, P., 1998, A\&AS 127, 277
\bibitem[1997]{kunzli} K\"unzli, M., North, P., Kurucz, R L. \& Nicolet, B.,
1997, A\&AS, 122, 51
\bibitem[1993]{kurucz1} Kurucz, R.L., 1993, ATLAS9 Stellar Atmosphere
  Programs and 2 km/s grid. 
\bibitem[1984]{kurucz} Kurucz, R.L., Furenlid, I., Brault, J. \& Testerman,
  L., 1984, \textit{The solar flux atlas from 296 to 1300 nm}, National Solar
  Observatory Atlas No. 1, Sunspot, New Mexico
\bibitem[1971]{maeder} Maeder, A., 1971, A\&A, 10, 354
\bibitem[1970]{maeder70} Maeder, A., Peytremann, E., 1970,A\&A, 7, 120
\bibitem[1991]{michaud} Michaud, G., 1991, in \textit{Evolution of
    Stars: the photoscpheric abundance connection}, IAU Symp 145,
  Eds. G. Michaud, and A. Tutukov, Kluwer, pp. 111
\bibitem[1983]{michaud83} Michaud, G., Tarasick, D., Charland,  Y.,
  Pelletier,  C., 1983, ApJ, 269, 239
\bibitem[1985]{MD85} Moon, T.T., Dworetsky, M.M., 1985, MNRAS, 217,305
\bibitem[1981]{N81} Nissen, P.E., 1981, A\&A, 97,145 
\bibitem[1997]{nissen97} Nissen, P.E., H\o g, E. \& Schuster, W.J., 1997
  in \textit{`Hipparcos - Venice '97'},  ESA SP-402, p. 225
\bibitem[1998]{north98} North, P., Erspamer, D. \& K\"unzli, M., 1998,
  contributions of the Astronomical Observatory Skalnate Pleso, 27, 252 
\bibitem[1997]{north} North, P., Jaschek, C. \& Egret, D., 1997, in:
  \textit{`Hipparcos - Venice '97'},  ESA SP-402, p. 367
\bibitem[2001]{queloz} Queloz, D., Mayor, M., Udry, S. et al., 2001,
  The Messenger, No. 105, p. 1 - 7 
\bibitem[1999]{queloz99} Queloz, D., Mayor, M., Weber, L. et al., 1999,
  A\&A, 354, 99
\bibitem[1996]{rentzsch} Rentzsch-Holm, I., 1996, A\&A, 312, 966
\bibitem[2001]{richard01} Richard, O., Michaud, G. \& Richer, J., 2001, ApJ,
  558, 377
\bibitem[2000]{richer00} Richer, J., Michaud, G. \& Turcotte, S., 2000, ApJ,
  529, 338
\bibitem[2002]{royer} Royer, F., Gerbaldi, M., Faraggiana, R. \& Gomez,
  A.E., 2002, A\&A, 381, 105
\bibitem[1988]{RN88} Rufener, F. \& Nicolet, B., 1988, A\&A, 206, 357
\bibitem[1992]{schaller} Schaller, G., Schaerer, D., Meynet, G. \& Maeder,
  A., 1992, A\&AS, 96, 269
\bibitem[1993a]{schaerer} Schaerer, D., Meynet, G., Maeder, A. \& Schaller,
  G., 1993a, A\&AS, 98, 523
\bibitem[1993b]{schaerer1} Schaerer, D., Charbonnel, C., Meynet, G.,
  Maeder, A. \& Schaller, G., 1993b, A\&AS, 102, 339
\bibitem[1997]{takeda97} Takeda, Y. \& Sadakane, K., 1997, PASJ, 49,367
\bibitem[1999]{thevenin99} Th\'evenin, F. \& Idiart, T.P., 1999, ApJ, 521, 753
\bibitem[1998]{turcotte98} Turcotte, S., Richer, J. \& Michaud, G., 1998,
  ApJ, 504, 559
\bibitem[1982]{uesugi} Uesugi, A. \& Fukuda, I., 1982, Catalogue of
  stellar rotational velocities (revised), Kyoto: University of Kyoto, 
  Department of Astronomy, 1982, Rev. Ed.
\bibitem[1999]{varenne99} Varenne, O. \& Monier, R., 1999, A\&A, 351, 247 

\end{thebibliography}
\end{document}